%% file: ProbingScrambling.tex
\documentclass[aps,prx,reprint,superscriptaddress,twocolumn]{revtex4-1}
\usepackage{graphicx}
\usepackage{amsmath}
\usepackage{amssymb}
\usepackage{braket}
\usepackage{color}

\definecolor{mygreen}{rgb}{0,0.5,0}
\definecolor{myblue}{rgb}{0,0,0.75}
\definecolor{mymagenta}{cmyk}{0,1,0,0.12}
\usepackage[colorlinks=true,
   linkcolor=blue,
   citecolor=blue,
   urlcolor =blue]{hyperref}

\newcommand{\citeSM}{\cite[{\tiny SM}\kern-0.3em][]{SM}}

\newcommand{\be}{\begin{equation}}
\newcommand{\ee}{\end{equation}}

\newcommand{\hilberth}{\mathcal{N}_\mathcal{H}}
\newcommand{\tr}{\mathrm{Tr}}
\expandafter\let\csname equation*\endcsname\relax
\expandafter\let\csname endequation*\endcsname\relax

\begin{document}
\title{Probing scrambling using statistical correlations between randomized measurements}

\author{B. Vermersch}
\email{benoit.vermersch@uibk.ac.at}
    \affiliation{Center for Quantum Physics,
University of Innsbruck, Innsbruck A-6020, Austria}
     \affiliation{Institute for Quantum Optics and Quantum Information, Austrian Academy of Sciences, Innsbruck A-6020,
      Austria}
    \author{A. Elben}
        \affiliation{Center for Quantum Physics,
University of Innsbruck, Innsbruck A-6020, Austria}
   \affiliation{Institute for Quantum Optics and Quantum Information, Austrian Academy of Sciences, Innsbruck A-6020,
      Austria}
        \author{L. M. Sieberer}
            \affiliation{Center for Quantum Physics,
University of Innsbruck, Innsbruck A-6020, Austria}
      \affiliation{Institute for Quantum Optics and Quantum Information, Austrian Academy of Sciences, Innsbruck A-6020,
      Austria}
    \author{N. Y. Yao}
    \affiliation{Department of Physics, University of California Berkeley, CA 94720, USA}
    \affiliation{Materials Science Division, Lawrence Berkeley National Laboratory, Berkeley, CA 94720, USA}
\author{P. Zoller}
        \affiliation{Center for Quantum Physics,
University of Innsbruck, Innsbruck A-6020, Austria}
    \affiliation{Institute for Quantum Optics and Quantum Information, Austrian Academy of Sciences, Innsbruck A-6020,
      Austria}
    
\begin{abstract}
We propose and analyze a protocol to study quantum information scrambling using statistical correlations between measurements, which are performed after evolving a quantum system from randomized initial states. We prove that the resulting correlations precisely capture the so-called out-of-time-ordered correlators and can be used to probe chaos in strongly-interacting, many-body systems.  Our protocol requires neither reversing time evolution nor auxiliary degrees of freedom, and can be realized in state-of-the-art quantum simulation experiments.
\end{abstract}
\maketitle

\section{Introduction}
Recent developments in quantum simulation have enabled the remarkable ability to interrogate and control atomic, molecular and ionic degrees of freedom in lattice experiments with single-site resolution~\cite{Gross2017,Blatt2012,Browaeys2016,Gambetta2017}.
In atomic Hubbard models with bosonic and fermionic atoms in optical lattices, a quantum gas microscope provides us with single-shot spatial- and spin-resolved images of atomic densities.  By averaging over many images, this allows one to extract spatial and spin equal-time correlation functions, which reveal unique properties of (non$-$)equilibrium quantum phases~\cite{Greif2016,Cheuk2016,Boll2016,Choi2016}.
In spin models, as realized with trapped ions~\cite{Zhang2017,Brydges2018}, Rydberg atoms~\cite{Zeiher2016,Guardado-Sanchez2018,Barredo2018,Keesling2018} and superconducting qubits~\cite{Blumoff2016,Barends2016,Otterbach2017,Gong2018}, the state of the spins (qubits) can be measured in a given standard basis in single-shot measurements with high fidelity and with high repetition rates.
Building on these  tools, we will develop below quantum protocols to measure many-body observables from analyzing \textit{statistical cross-correlations} between such quantum images representing different runs of an experiment.
We will see that this provides us with simple, generic and robust techniques to extract many-body observables, which are challenging to access otherwise within existing experimental setups.
In particular, we will develop novel protocols for out-of-time-ordered correlators (OTOCs), which are time-dependent quantities which cannot be measured directly as a standard time-ordered correlation function.
OTOCs represent a key quantity to diagnose quantum chaos and enable one to understand how quantum information propagates, and ``scrambles''~\cite{Hosur2016}, in close connection to the notion of entanglement spreading~\cite{Calabrese2016,Bohrdt2017,VonKeyserlingk2018, Nahum2018}.

OTOCs have been introduced to characterize quantum dynamics, described by a unitary time evolution operator $U(t)$ in terms of the complexity of Heisenberg operators $W(t) = U^{\dagger}(t) W U(t)$.
For chaotic dynamics, even an initially ``simple'' and local Hermitian operator $W$ rapidly becomes
complex and non-local. 
As a consequence, after a short time, $W(t)$ is
delocalized and no longer commutes with an initially non-overlapping local operator $V$.
The degree of non-commutativity, or equivalently the scrambling of $W(t)$ is quantified by the out-of-time-ordered correlator (OTOC), which takes the form
\begin{eqnarray}\label{eq:OTOC}
O(t) = \tr( \rho W(t) V^\dag  W(t) V )/ \tr( \rho W(t)^2 V^\dag V ),
\end{eqnarray}
with $\rho$ the initial quantum state. Note that this definition ensures that $O(t)=1$ when $W(t)$ and $V$ commute.
 In the following, we focus on the
``infinite-temperature'' OTOC~\cite{Fan2016, Chen2017, VonKeyserlingk2018,
  Nahum2018} for which $\rho= I/\mathcal{N_H}$, with $I$ the identity operator
and $\mathcal{N_H}$ the Hilbert space dimension.
We will discuss in the outlook extensions of our approach to thermal states.
The time dependence of $O(t)$ can distinguish between different classes of scrambling,
ranging from ``fast scrambling''  in models with holographic duals~\cite{Sachdev1993,
  Kitaev, Hayden2007, Shenker2014, Maldacena2016, Banerjee2017} and chaotic
many-body spin systems~\cite{Hosur2016,Roberts2016, VonKeyserlingk2018, Nahum2018,Pappalardi2018}, to ``slow scrambling" characteristic of many-body localization (MBL)~\cite{Fan2016, Chen2017,Syzranov2017}.
These theoretical insights raise the question of how to experimentally measure $O(t)$, despite the peculiar time order inherent in its definition.
A first option to measure $O(t)$ consists in implementing time-reversal operations~\cite{Swingle2016, Zhu2016, Bohrdt2017,Shen2017, Yoshida2018}, or using auxiliary quantum systems~\cite{Yao2016}.
The first measurements of OTOCs were realized using this approach in systems with few degrees of freedoms~\cite{Li2017}, but also in a trapped ion setup with infinite-range interactions~\cite{Garttner2017}. 
However, protocols based on realizing time-reversal operations remain for many experimental platforms --- like Hubbard systems or systems with local interactions --- an experimental challenge. For such protocols, recent studies have also shown that decoherence can ``mimick'' the effect of scrambling, and developed~\cite{Yoshida2018,Swingle2018} and realized~\cite{Landsman2018} implementations involving auxiliary degrees of freedoms to distinguish the two effects.
\begin{figure*}
  \includegraphics[width=\textwidth]{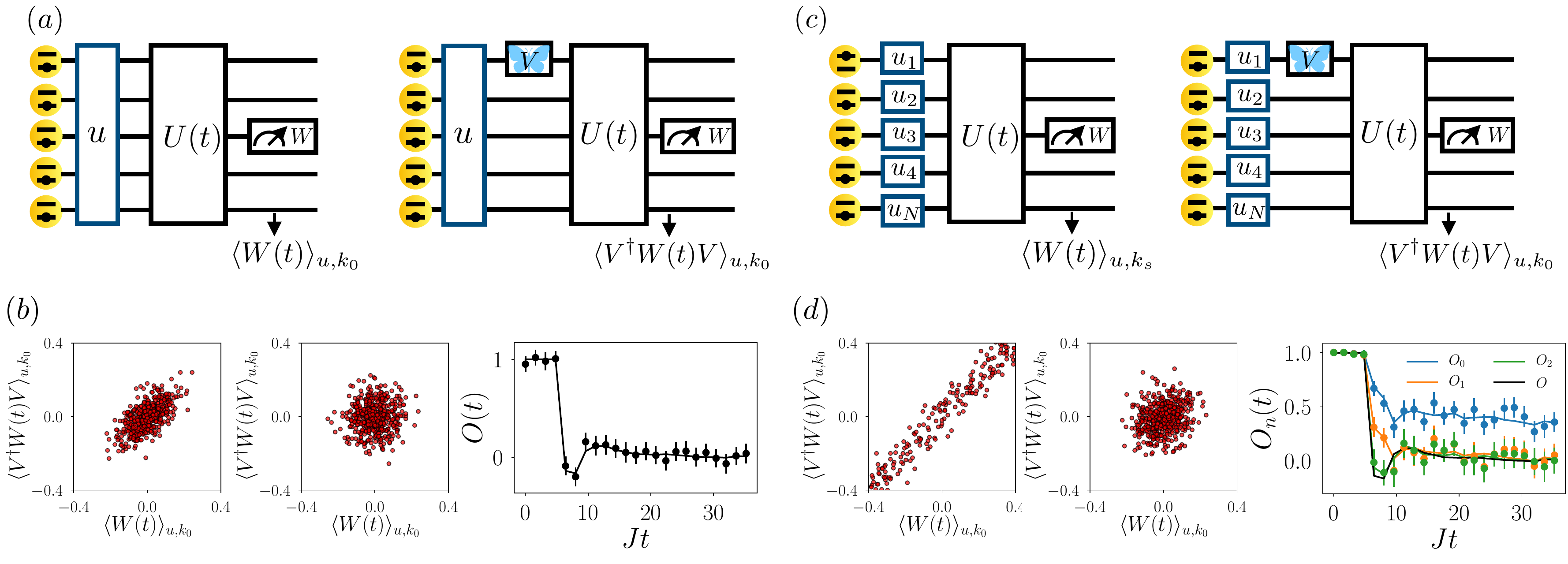}
  \caption{{\it Probing scrambling via statistical correlations in a spin system.}
     (a) The global protocol consists in measuring separately
    $\langle W(t) \rangle_{u,k_0}$ and $\langle V^\dagger W(t) V \rangle_{u,k_0}$ to obtain the OTOC $O(t)$. 
    (b) Numerical simulations of the protocol for the kicked Ising model with parameters $h_x=J$, $h_z=0.809J$, $JT=1.6$, $N=8$, and $j=3$.
    The first two panels show the statistics of the measurement at $Jt=0$, and $Jt=35.2$
    In the third panel, the exact OTOC $O(t)$ is shown as solid line, and the points represent the statistical correlations obtained from $N_u=500$ unitaries and $N_M=500$ projective measurements (per unitary), with errors bars placed at $\pm 2 \sigma$ calculated from the Jacknife resampling method.
    (c) Local protocol using local random unitaries.
    (d)   Same as (b) with the local protocol, extracting the modified OTOC $O_n(t)$ (calculated exactly, colored lines) from the statistical correlations (circles with error bars). Convergence to the OTOC $O(t)$ (black line) is achieved for low index $n\sim2$. Throughout this work, we use unbiased estimators for the normalization terms $\mathcal{D}^\mathrm{(G,L)}$. 
    \label{fig:setup}}
\end{figure*}

In contrast, a unique feature of our protocols to measure OTOCs via statistical correlations is that they do not rely on time-reversal operations nor the presence of an ancillary system. 
In addition, we will show that OTOCs extracted from statistical averages provide the key advantage to be naturally robust against various forms of decoherence and experimental noise,  including depolarization and readout errors.
As a consequence, our protocols can be realized in any state-of-the-art AMO~\cite{Bloch2012,Blatt2012,Browaeys2016} or superconducting qubit platforms~\cite{Gambetta2017}, and used as experimental probes of scrambling in many-body systems.

The present paper presents two key results related to two protocols. First, we present the \emph{global protocol} and demonstrate the exact equivalence between the OTOC $O(t)$, as defined above, and the statistical correlations obtained from initial states, which are randomized with a \emph{global} unitary operator $u$ for the total many-body system.
We then present the \emph{local protocol}, which consists of an experimentally simpler approach for spin systems, where the initial state is randomized with \emph{local} unitary operations, and where the statistical correlations also give access to  $O(t)$.

This paper is organized as follows. Section \ref{sec:protocols} presents the main results of this paper by describing the two protocols to measure $O(t)$ with local and random unitaries.
Section~\ref{sec:examples} gives different physical examples, accessible to current AMO and solid-state platforms. Finally, we discuss in section \ref{sec:errors} the role of statistical errors and imperfections, and identify in particular the different situations (depolarization, readout errors) where the protocol is not affected.

\section{Protocols mapping statistical correlations to OTOCs}\label{sec:protocols}

In this section, we present and illustrate both the \emph{global} and \emph{local} protocols  to measure OTOCs via random measurements.
We consider a system $\mathcal{S}$ associated with a Hilbert space of dimension $\mathcal{N_H}$.
This can be, for example, a set of atoms described by a Hubbard model, or an ensemble of spin-$1/2$ as shown in Fig.~\ref{fig:setup}(a).
In the following, we also assume the operator $V$ to be unitary, and the operator $W$ to be Hermitian and traceless ($\mathrm{Tr}(W)=0$).  Note that these conditions do not restrict the ability of OTOCs to describe scrambling: For spin systems we will consider as examples $W$ and $V$ to be local Pauli operators, which are particularly relevant in this context~\cite{VonKeyserlingk2018, Nahum2018}. We also give examples below that are relevant to probe scrambling in Hubbard systems.

Our two protocols are illustrated in Fig.~\ref{fig:setup}. 
The first protocol described in Sec.~\ref{sec:global} relies on \emph{global} random unitaries $u$ from the circular unitary ensemble CUE($\mathcal{N_H}$)~\cite{Haake2010} or from a unitary 2-design~\cite{Dankert2009}. As illustrated below for a Bose-Hubbard systems, such random unitaries can be realized in generic interacting models using time-dependent disorder~\cite{Nakata2017,Elben2018,Vermersch2018}.
The second protocol presented in Sec.~\ref{sec:local} considers spin systems with individual spin control, which allows us to simplify drastically the experimental task by replacing the global random unitaries by \emph{local} random unitaries $u=u_1\otimes u_2\otimes \ldots \otimes u_N$, $u_i \in \mathrm{CUE}(2)$, acting on single spins $i=1,\dots,N$. 
Note that such local random unitaries have been recently realized with high fidelity with trapped ions~\cite{Brydges2018,Elben2018a}.

 We find it convenient to present both protocols as experimental recipes to measure the statistical correlations. In each case, we then relate mathematically the correlations to $O(t)$.

\subsection{The global protocol}\label{sec:global}

{\it Experimental protocol.---} The protocol consists of the following steps, as illustrated in Fig.~\ref{fig:setup}(a).

(i) We prepare an arbitrary state $\ket{k_0}$, and apply a {\em global} random unitary $u$ to obtain $\ket{\psi_{u,k_0}}=u\ket{k_0}$.
The randomized state $\ket{\psi_{u,k_0}}$ is the starting point for two independent experiments: 

(ii.a) In the first experiment, we evolve
the system in time with $U(t)$, and perform a measurement of the expectation value of $W$.
The time evolution operator $U(t)$ can be generated for instance from a static Hamiltonian $U(t)=e^{-iHt}$, from Floquet evolution, or from a quantum circuit operating on qubits, depending on the type of system under study.
We repeat steps
(i) and (ii.a) with the same random unitary $u$ to measure 
$\langle W(t) \rangle_{u,k_0} = \bra{\psi_{u,k_0}} W(t) \ket{\psi_{u,k_0}}$, as illustrated in
Fig.~\ref{fig:setup}(a). 

(ii.b) In the second
experiment, we prepare again $\ket{\psi_{u,k_0}}$ and apply the unitary $V$. This operation is followed by the time evolution with $U(t)$, and
a measurement of $W$. We repeat this sequence to obtain
$\langle V^\dagger W(t) V \rangle_{u,k_0} = \bra{\psi_{u,k_0}}  V^\dagger W(t) V \ket{\psi_{u,k_0}}$, as shown in Fig.~\ref{fig:setup}(a). 

(iii) Finally, we repeat steps (i) and (ii) for different random unitaries. The OTOC $O(t)$, as defined in Eq.~\eqref{eq:OTOC}, is then obtained from the statistical correlations
\begin{equation}
  \label{eq:identityOG}
  O(t) = \frac{1}{\mathcal{D}^\mathrm{(G)}}\overline{\langle W(t) \rangle_{u,k_0} \langle V^\dag W(t) V  \rangle_{u,k_0}},
\end{equation} 
between the measurement outcomes
$\langle W(t) \rangle_{u,k_0}$ and $\langle V^\dagger W(t) V \rangle_{u,k_0}$ of (ii.a) and (ii.b), respectively. Here, $\overline{\vphantom{V }\cdots}\ $
denotes the ensemble average over random unitaries $u$, and $\mathcal{D}^\mathrm{(G)}={\overline{\langle W(t) \rangle_{u,k_0} ^2 }}$ is a normalization term.

\begin{figure}
\includegraphics[width=0.85\columnwidth]{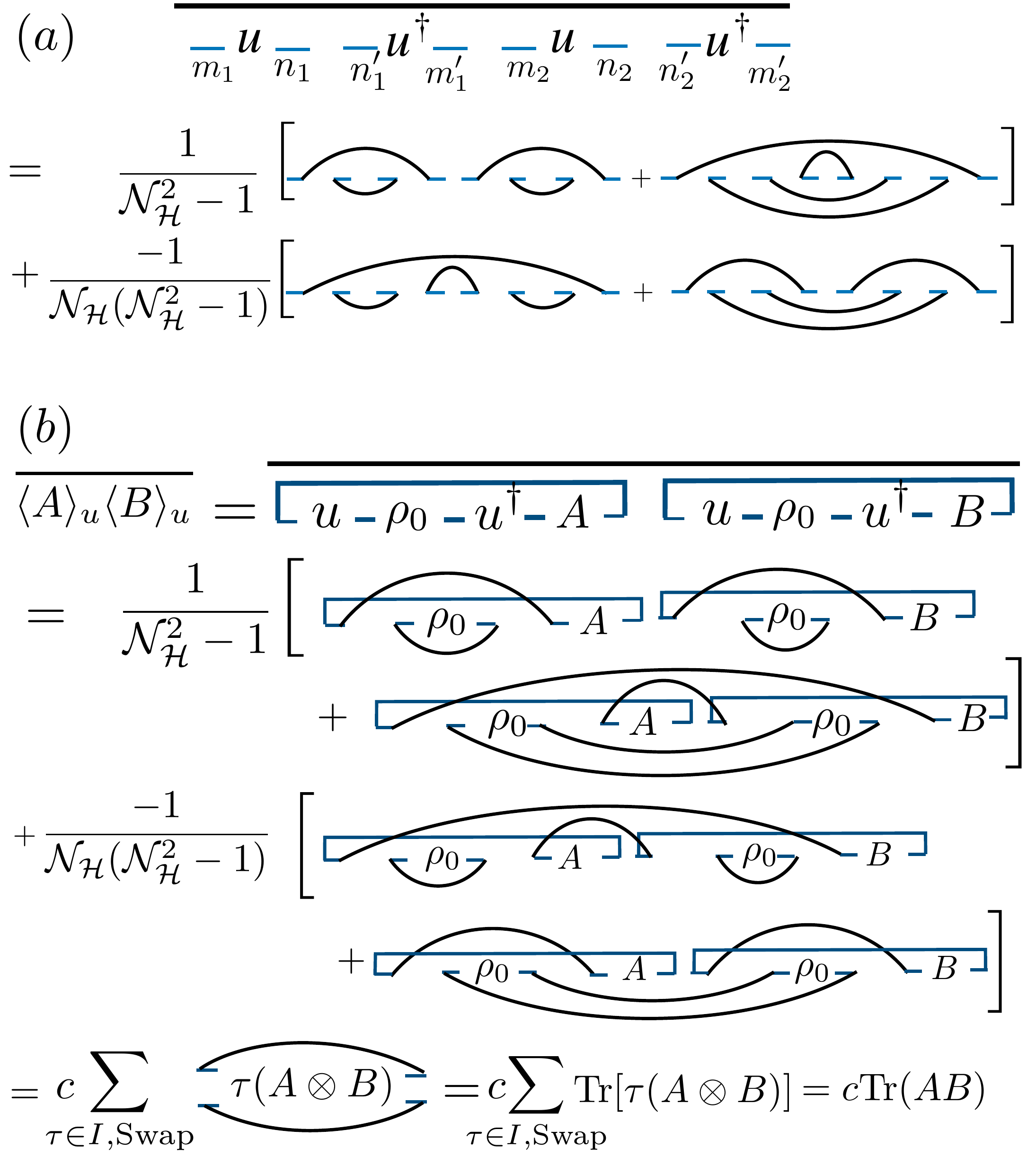}
\caption{{\it Identities for our protocol with global unitaries.} (a) Diagrammatic representation of the 2-design identities of the CUE Eq.~\eqref{eq:2design}. (b)  Correlations between two measurements with $A=W(t)$, $B=V^\dagger W(t) V$.
Here we have introduced the
notation for the initial pure density matrix
$\rho_0=\ket{k_0}\bra{k_0}$. \label{fig:Proof_G}}
\end{figure}

{\it Proof and illustration.---}
Eq.~\eqref{eq:identityOG} can be proven using the 2-design identities, which provide analytical expressions for the statistical correlations between the matrix elements of $u$~\cite{Collins2006},
\begin{eqnarray}
&&\hspace{2cm}\overline{u_{m_1,n_1} u_{m'_1,n'_1}^*u_{m_2,n_2}u_{m_2',n_2'}^* } \label{eq:2design} 
 \\
&=& \frac{\delta_{m_1,m_1'}\delta_{m_2,m_2'}\delta_{n_1,n_1'}\delta_{n_2,n_2'}+\delta_{m_1,m_2'}\delta_{m_2,m_1'}\delta_{n_1,n_2'}\delta_{n_2,n_1'}}{\hilberth^2-1}
\nonumber \\
&-&\frac{\delta_{m_1,m_1'}\delta_{m_2,m_2'}\delta_{n_1,n_2'}\delta_{n_2,n_1'}+\delta_{m_1,m_2'}\delta_{m_2,m_1'}\delta_{n_1,n_1'}\delta_{n_2,n_2'}}{\hilberth(\hilberth^2-1)}, \nonumber 
\end{eqnarray}
with $\delta$ the Kronecker delta. In order to simplify the proofs, we use in this work a diagrammatic representation of
Eq.~\eqref{eq:2design} where the contraction of the different
indices are represented by lines~\cite{Brouwer1996}, as shown in Fig~\ref{fig:Proof_G}(a). 
This allows us to prove in Fig.~\ref{fig:Proof_G}(b) the identity
\begin{eqnarray}
\overline{\langle W(t) \rangle_u \langle V^\dagger W(t) V \rangle_u}
&=&c\! \sum_{\tau =I,\mathrm{Swap}}\! \mathrm{Tr}[\tau (W(t)\otimes V^\dagger W(t) V)]\nonumber \\ 
&=& c\ \tr [W(t)V^\dagger W(t) V],\label{eq:WVWV}
\end{eqnarray}
with $c=[\mathcal{N}_\mathcal{H}(\mathcal{N}_\mathcal{H}+1)]^{-1}$. 
The trace in the first line is performed over an extended Hilbert space $\mathcal{H}\otimes \mathcal{H}$, where $\mathcal{H}$ is the Hilbert space of dimension $\mathcal{N}_\mathcal{H}$, and the swap operator is $\mathrm{Swap}(\ket{k}\otimes\ket{k'})=\ket{k'}\otimes\ket{k}$ for each pair of states $\ket{k}$, $\ket{k'}$. 
The condition of traceless operator $\mathrm{Tr}[W(t)]=0$, and the identity $\mathrm{Tr}[\mathrm{Swap}(W(t)\otimes V^\dagger W(t) V)]=\mathrm{Tr}[W(t)V^\dagger W(t) V]$ yield the second line of Eq.~\eqref{eq:WVWV}. Finally, to conclude our proof, we use the same identity with $V\to I$ to prove that the denominator in Eq.~\eqref{eq:identityOG} reduces to $c\mathrm{Tr}(W^2(t))$. 

As an illustration, we present in
Fig.~\ref{fig:setup}(b) the intuitive physical picture behind this result (based on realizing $U(t)$ for the kicked Ising model, see caption and text below).  
At $t=0$, the measurement of $W(0)=W = \sigma^z_j$ ($j>1$) is not affected by the operator $V = \sigma^z_1$, which distinguishes the two initial states $\ket{\psi_{u,k_0}}$ and $V\ket{\psi_{u,k_0}}$. 
 Indeed, $[V, W(0)] = 0$ and hence
$\langle W(0) \rangle_{u,k_0} = \langle V^\dagger W(0) V \rangle_{u,k_0}$, implying perfect
correlations, i.e., $O(0)=1$ (up to shot noise errors, see below), see Fig.~\ref{fig:setup}(b) left panel. 
At later times (middle panel), when $[V, W(t)] \neq 0$ due to the spreading of $W(t)$, the value of
$\langle V^\dagger W(t) V\rangle_{u,k_0}$ becomes decorrelated from $\langle W(t)\rangle_{u,k_0}$. Note that in analogy to our approach the distance between a quantum state and a (physical) copy, which is perturbed at $t=0$ by the operator $V$,  has been proposed to numerically detect scrambling~\cite{Leviatan2017,Chen2017a}.

In an experiment, a finite number of $N_u$ random unitaries is realized to measure $O(t)$ based on Eq.~\eqref{eq:identityOG}. 
Furthermore the operator $W$ is measured via a finite number of projective measurements $N_M$ per unitary. 
The finite values of $N_u$ and $N_M$ will thus lead to statistical errors.
To illustrate this aspect, we  compare in Fig.~\ref{fig:setup}(b) the time evolution of $O(t)$ with the estimation obtained from a finite realistic number of measurements~\cite{Brydges2018} (circles with statistical error bars). We analyze in more details in Sec.~\ref{sec:errors} the role of statistical errors. 

One of the experimental challenges in the protocol presented above consists in generating, with high fidelity, {\em global} random unitaries $u$ satisfying the $2-$design properties. In quantum simulators implementing Hubbard or spin models, this can be done using random quenches based on time-dependent disorder potentials (see Refs.~\cite{Elben2018,Vermersch2018} and example in Sec.~\ref{sec:BH}).
We now proceed to describe an experimentally significantly simpler protocol for spin systems, which only requires to generate {\em  local} random unitaries acting on individual spins.
These unitaries can be realized by combining local rotations along a fixed axis of the Bloch sphere, say the $z$ axis, with global rotations along an orthogonal direction, for instance the $x$ axis~\cite{Brydges2018,Elben2018a}, and can therefore be implemented in present qubit experiments with single-site control, e.g.~with trapped ions~\cite{Blatt2012}, Rydberg atoms~\cite{Browaeys2016} or superconducting qubits~\cite{Gambetta2017}. 

\subsection{The local protocol}\label{sec:local}

{\it The protocol---}
We now describe our protocol based on local unitaries. 
The main difference compared to the protocol presented in Sec.~\ref{sec:global} is that we need to consider here an ensemble $E_n=\{\ket{k_0},\dots\}$ of initial states, instead of a single one $\ket{k_0}$, in order to obtain a mapping between statistical correlations and OTOCs. The states $\ket{k_s}=\ket{k_s^{(1)},k_s^{(2)},\dots}$, which we consider are written as product states in a standard fixed basis, i.e., $k_s^{(i)}=\uparrow,\downarrow$, and could be easily prepared in an experiment with single-site control.
For this second protocol,  each state $\ket{k_s}\in E_n$ is subject to the same local unitary $u=u_1\otimes\dots u_N$, which gives access to the random measurement $\langle W(t) \rangle _{u_,k_s}$ after time-evolution with $U(t)$, c.f. Fig.~\ref{fig:setup}(c). In a second step, we access $\langle V^\dag W(t) V\rangle _{u_,k_0}$ for a single initial state $\ket{k_0}$, e.g., $\ket{k_0}=\ket{\downarrow,\dots,\downarrow}$, and the same random unitary $u$, as in Sec.~\ref{sec:global}. 
From these measurements, we can construct the statistical correlations
 \begin{eqnarray}
\label{eq:defOLn}
O_n(t) 
&=&   \frac{1}{\mathcal{D}_n^\mathrm{(L)}}\sum_{k_s\in E_n}  c_{k_s} \overline{\langle W(t) \rangle_{u,k_s} \langle V^\dag  W(t) V  \rangle_{u,k_0}}, 
\end{eqnarray}
with \mbox{$\mathcal{D}_n^\mathrm{(L)}= \sum_{k_s\in E_n} c_{k_s}\overline{\langle W(t) \rangle_{u,k_s} \langle W(t)   \rangle_{u,k_0}}$}, and weights $c_{k_s}$.
In the following, we show how to choose the ensembles $E_n$ and weights $c_{k_s}$, so that the correlations $O_n(t)$ represent a converging series, indexed by $n$, of ``modified OTOCs" approximating $O(t)$. 
The low-order OTOCs $O_{0,1,\dots }(t)$, which correspond to small numbers of initial states to sample, and are thus the easiest quantities to access experimentally, provide generically good approximations of $O(t)$. We will also prove, in the other limit $n=N$,  the exact relation $O_N(t)=O(t)$.

{\it Introducing the modified OTOCs---}
Here, for concreteness we consider $V$ to be a Pauli operator on the first site $i=1$.
For each value of $n=0,\dots,N$, we then define the ensemble $E_n$ as the set of all $2^n$ configurations $\ket{k_{s}}$, such that only the states of the first $n$ spins can differ from the ones of the reference state $\ket{k_0}$, i.e., $k^{(i>n)}_s=k^{(i)}_0$ for $i>n$. For instance, for $n=0$ ($n=1$, respectively), which  we study in detail below, the ensemble $S_0=\{\ket{k_0}\}$ ($S_1=\{\ket{k_0},\sigma_1^x\ket{k_0}\}$) is represented by a single state (just two states). 
As proven in App.~\ref{app:local}, by choosing the weights $c_{k_s}=(-1/2)^{d[k_0,k_s]}$, with $d[k_0,k_s]$ the Hamming distance (the number of spin flips between $\ket{k_0}$ and $\ket{k_s}$), we can relate the statistical correlations $O_n(t)$ to  ``modified OTOCs"
\begin{eqnarray}
\label{eq:identityOLn}
 &&O_n(t)=\frac{\sum_{A, B_n \subseteq A} \mathrm{Tr}_{A}\left(W(t)_A (VW(t)V)_A\right)}
 {\sum_{A, B_n \subseteq A} \mathrm{Tr}_{A}\left(W(t)_A W(t)_A\right)},
\end{eqnarray}
which converge to $O(t)$ for $n=N$~\footnote{We use here $\mathrm{Tr}(W(t)^2)=2^N$.}. Here, the sums in the first line are performed over all partitions $A$ which include the set $B_n=\{1,\dots,n\}$ of the first $n$ spins (for $n=0$, $B_0=\emptyset$ is empty), and the traces are performed over 
reduced operators $W_A(t)=\mathrm{Tr}_{\mathcal{S}-A}(W(t))$, and $(VW(t)V)_A=\mathrm{Tr}_{\mathcal{S}-A}(VW(t)V$). 
The modified OTOCs $O_n(t)$ are thus sums of out-of-time-ordered functions of the different reduced operators $W(t)_A$, $(VW(t)V)_A$. 

{\it Properties of the modified OTOCs and illustrations---}
The identity Eq.~\eqref{eq:identityOLn} shows that the index $n$ plays the role of a spatial resolution controlling how $O_n(t)$ approximates $O(t)$. For all contributing partitions $A$,  $\{1\dots n\} \subseteq A$, the information about the first $n$ spins is preserved when reducing the operators $W(t)\to W(t)_A$ $VW(t)V\to (VW(t)V)_A$.
In particular, for the maximal spatial resolution $n=N$, statistical correlations are exactly the OTOC $O_N(t)=O(t)$. In the opposite case of $n=0$, all partitions $A\subseteq \mathcal{S}$ of the system contribute to $O_0(t)$. For $O_1(t)$, the information related to the support of $V$ (here the first site) is ``resolved'' so that we can expect a better approximation to $O(t)$, and so on for $n=2,3\dots$.
Note that our construction of the sets $E_n$ can be generalized easily to other positions of $V$, but also to multi-site operators.

An illustration of this protocol is shown in Fig.~\ref{fig:setup}(d) [compare to Fig.~\ref{fig:setup}(b)], where we represent $O_{n}(t)$ (solid lines), and the corresponding statistical correlations obtained by simulating numerically the protocol. For $n=0$, the modified OTOC captures the scrambling time as $O(t)$ but saturates to a non-zero value at long times. For $n=1,2,3\dots$  the values of $O_{n}(t)$ are in good quantitative agreement with $O(t)$. We also note that for short times the local protocol has an advantage compared to the global protocol in terms of statistical errors, as we explain below.

\section{Realizations of the protocols in different physical scenarios}\label{sec:examples}
This section is devoted to physical examples which can be accessed with our protocol. In this first case Sec.~\ref{sec:BH}, we show how to apply the global protocol in an atomic Bose-Hubbard system. We then focus on the local protocol and analyze for chaotic (Sec.~\ref{sec:chaos}), many-body localized (Sec.~\ref{sec:MBL}) and long-range spin models (Sec.~\ref{sec:Longrange}) the behavior of modified OTOCs $O_n(t)$. We analyze in particular both via  analytical models and numerical simulations the convergence properties of $O_n(t)$ to $O(t)$.

\subsection{Implementation of the global protocol in a Bose-Hubbard chain}\label{sec:BH}
We now present an example to illustrate the different steps of the protocol with global unitaries and consider the situation of scrambling dynamics of the Bose-Hubbard (BH) chain~\cite{Bohrdt2017}, with $U(t)=\exp(-iH_\mathrm{BH} t)$, 
\begin{equation}
H_\mathrm{BH} = -J \sum_{i=1}^N \left(a^\dag_{i+1}a_i+\mathrm{h.c.}\right)  + \frac{U_\mathrm{int}}{2} \sum_{i=1}^N n_i (n_i-1), 
\end{equation}
with $a_i$ bosonic operators and $n_i=a^\dag_i a_i$. 
We consider here the OTOC dynamics for the unitary $V=\exp(-i \pi a_1^\dag a_1)$ and the traceless observable $W=n_{j+1} - n_{j}$.

\begin{figure}
\includegraphics[width=\columnwidth]{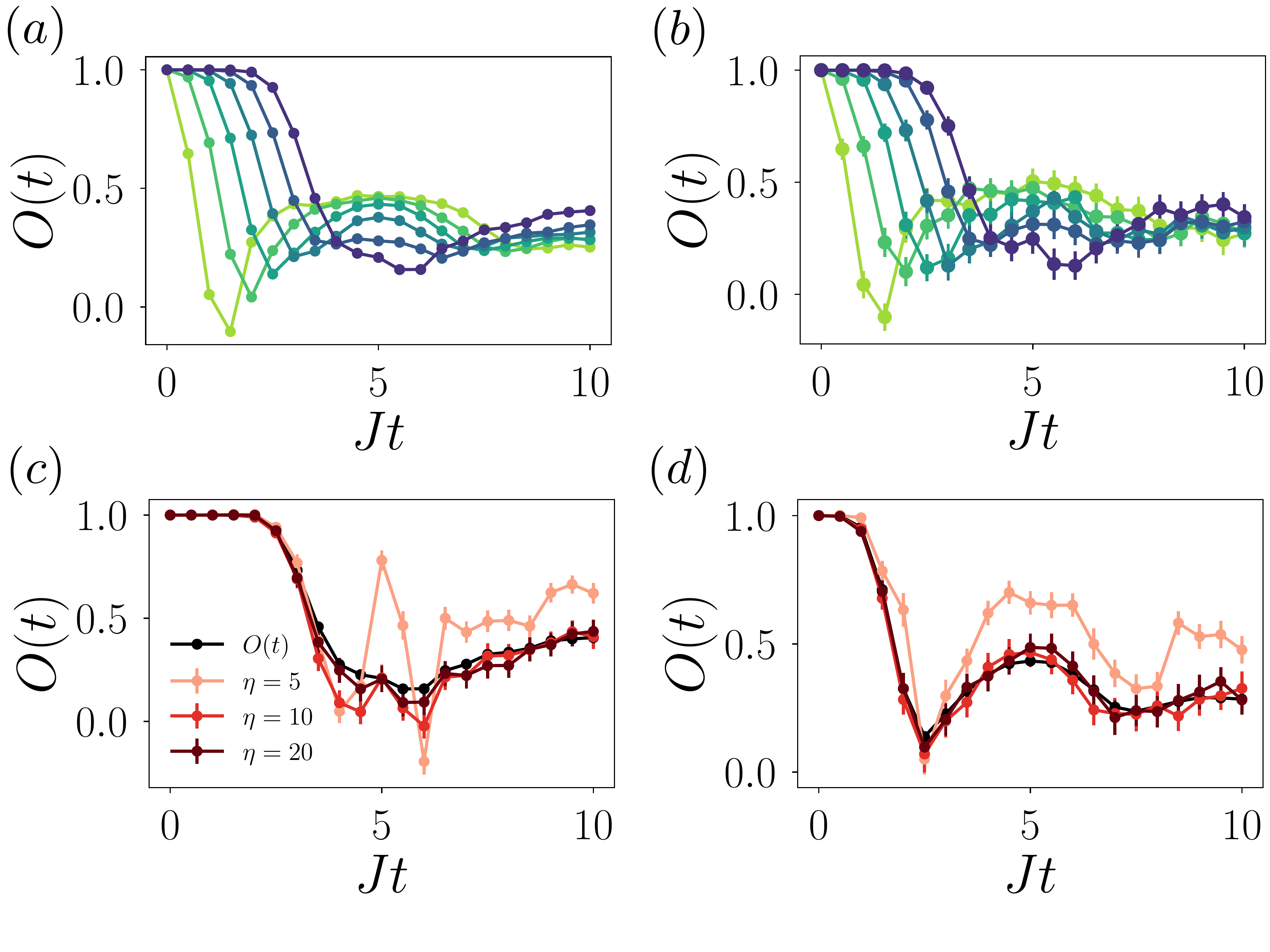}
\caption{{\it Scrambling in the BH chain}\label{fig:BH}
(a) OTOC $O(t)$ for different positions $j=2,\dots,N-1$ (green to purple) of the operator $W$ (see text).
(b) $O(t)$ estimated with $N_u=1000$ unitaries $u$ sampled numerically from the Haar measure.
(c-d) Comparison between $O(t)$ and corresponding estimations via statistical correlations, where $u$ is prepared via $\eta$ random quenches, with $j=7$ (c) and $j=4$ (d), and $N_u=1000$.
Here, $\Delta_j^{(m)}$ is sampled from a uniform distribution of width $2J$, and the quench time is $T=1/J$. For all panels, we consider $N=8$ lattice sites with $\ket{k_0}=\ket{10101010}$ in the number basis, and $U_\mathrm{int}=2J$.
In panels (b-d), error bars at $2$ standard deviations are calculated from Jacknife resampling method. Here, we consider $N_M\to\infty$.
}
\end{figure}

We first illustrate the mapping Eq.~\eqref{eq:identityOG} in Fig.~\ref{fig:BH} by comparing $O(t)$ [panel (a)] and the corresponding statistical correlations [panel (b)], where $N_u=1000$ random unitaries were sampled numerically from the CUE. Assuming here no projection noise $N_M\to \infty$, we obtain a very good agreement between the two quantities. Note as in the case of the spin models described below the characteristic ``scrambling time" for the OTOCs varies essentially linearly with the position of the operators, showing the existence of a ``butterfly" velocity.

We now discuss the physical realization of the global random unitaries $u$. Unitaries satisfying the required 2-design properties can be generated using the same Hamiltonian $H_\mathrm{BH}$ subject to a sequence of $\eta$ random quenches~\cite{Elben2018,Vermersch2018}
\begin{equation}
u = \prod_{m=1}^\eta \exp \left(-i T \left[H_\mathrm{BH}+\sum_j \Delta_j^{(m)} n_j \right]\right),
\end{equation}
with $\Delta_j^{(m)}$ a random disorder potential, which is reinitialized for each quench $m$, and the quench time $T$. As shown in Refs.~\cite{Elben2018,Vermersch2018}, such  random quenches generate efficiently, in each particle number sector, unitaries satisfying the required $2$-design properties of the CUE, after a time $\eta T\approx N$. This is illustrated in Fig.~\ref{fig:BH}(c-d) for two different operators $W$ positions:  for $\eta\ge N$, the generated $u$ converged to the CUE (with respect to the required $2$-design properties), and therefore the measured statistical correlations coincide within the statistical error bars with $O(t)$. 

\subsection{Chaotic dynamics with modified OTOCs}\label{sec:chaos}
In the rest of this section, we focus on spin models with OTOCs measured by local unitaries. 
To illustrate the ability of the modified OTOCs $O_n(t)$ Eq.~\eqref{eq:identityOLn} to approximate $O(t)$, we first consider scrambling in chaotic spin models, which is characterized by two key features:
The support of an operator
$W(t)$ that is initially localized grows ballistically with a ``butterfly velocity'' $v_B$, and the operator front
traveling at $v_B$ broadens  diffusively~\cite{Hosur2016, VonKeyserlingk2018,Nahum2018}. 
The ballistic growth can be captured by a simple phenomenological
model, which assumes that $U(t)$ takes the form \mbox{$
  U(t) = U[L(t)]_1 \otimes \dots\otimes U[L(t)]_{N/L(t)}$}, 
where the Haar random unitaries $U[L(t)]\in \mathrm{CUE}(2^{L(t)})$ describe
scrambling on a linearly growing scale $L(t)=1+\mathrm{floor}(v_B t)$. For $W=\sigma_j^z$ and $V=\sigma_1^z$, we obtain in leading order in $L(t)\gg1$~\cite{Roberts2016}
\begin{eqnarray}
O(t)&=&1 \hspace{0.2cm} (j>L(t))  \hspace{0.3 cm}O(t)_{\phantom{0}}=-\frac{1}{4^{L(t)}}   \hspace{0.3cm} (j\le L(t)), \label{eq:scrambling1}
\end{eqnarray}
and, as shown in App.~\ref{app:haar},
\begin{eqnarray}
O_0(t)&=&1 \hspace{0.1cm} (j>L(t))  \hspace{0.15 cm}O_0(t)=\frac{1}{3}    \hspace{1.2cm} (j\le L(t)) \label{eq:scrambling2} \\
O_1(t)&=&1 \hspace{0.1cm} (j>L(t))  \hspace{0.15 cm}O_1(t)= -\frac{1}{2^{L(t)+1}}   \hspace{0.05cm} (j\le L(t)), \nonumber \label{eq:scrambling3}
\end{eqnarray}
Here, $O(t)$ represents the average OTOC over the unitaries $U$, whereas the expression for $O_{n}(t)$ corresponds to including the sampling over $U$ in the ensemble average $\overline{\vphantom{V }\cdots}\ $~\footnote{For all numerical simulations presented in this work, we consider $N_u$ independent samplings of the unitaries $u$ and $U$, i.e we sample new pairs of unitaries $u$ and $U$ for each realization}. All OTOCs $O(t)=O_\mathrm{0,1}(t)=1$ coincide thus at
short times when $W(t)$ and $V$ commute exactly. At the ``scrambling time'' $t_B = r/v_B$ ($r=j-1$), $O(t)$ and $O_\mathrm{0,1}(t)$ exhibit a sharp drop. For $t>t_B$,
$O(t)$, and $O_1(t)$ are exponentially suppressed while $O_\mathrm{0}(t)$ converges to $1/3$.
\begin{figure}
  \includegraphics[width=\columnwidth]{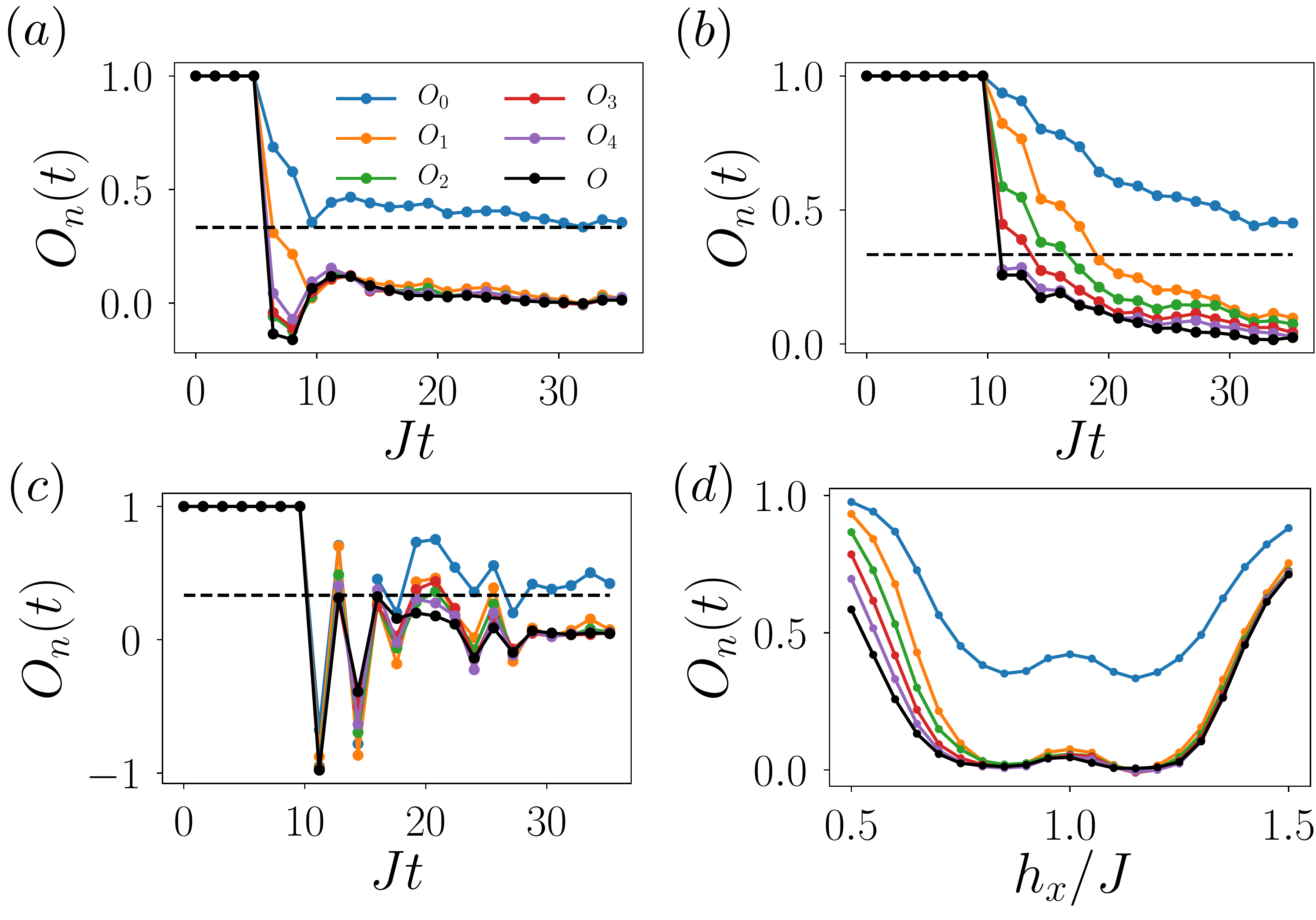}
  \caption{ {\it $O_n(t)$ vs.~$O(t)$ for chaotic dynamics in the kicked Ising model.}
    All plots represent for $N=8$ sites the modified OTOCs $O_n(t)$, for $n=0,1,2,3,4$ and the OTOC $O(t)$.
    The dashed line shows the predicted value $1/3$ for $O_0(t)$ for Haar scrambling.
    (a) For $h_x=0.75J$, and $j=3$, the OTOC decay is fast and well captured by the low-order OTOCs $O_n(t)$.
    (b) For a larger $j=7$ operator position, the OTOC decay is slower, as a consequence of the diffusive broadening of $W(t)$.
    (c) For $j=7$, and $h_x=J$, the decay of $O(t)$ is fast but accompanied by strong oscillations, which are also resolved by the modified OTOCs.
    (d) For a fixed long time $Jt=35.2$, the modified OTOCs provide good approximations to $O(t)$, in particular in the regime $h_x\ge0.75J$.
    Other parameters are $h_z=0.809J$, and $JT=1.6$.
    \label{fig:scrambling}}
\end{figure}

We now confront these analytical predictions with numerical simulations of the kicked Ising model, which is an example of a chaotic spin model~\cite{VonKeyserlingk2018}
\begin{eqnarray}
U(mT)= \left [e^{-i\frac{T}{2} (\sum_i J \sigma^z_i \sigma^z_{i+1}+h_z  \sigma_i^z)  } e^{-i\frac{T}{2} h_x \sum_i  \sigma_i^x }\right]^m, 
\end{eqnarray}
with $m$ a positive integer, and $T$ the period of the Floquet system. Throughout this work, we use open boundary conditions (OBC). 
The results are shown in Fig.~\ref{fig:scrambling}, where we compare the modified OTOCs for $n=0,1,2,3,4$ with $O(t)$. 
In panel (a) corresponding to an operator position $j=4$, the scrambling dynamics described by $O(t)$ is fast in the sense that it occurs at a time $t_B\sim j/J$. This behavior is qualitatively captured by the first modified OTOC, with fast decay at the scrambling time and saturation at the predicted value $1/3$ of our phenomenological model.
Interestingly, the convergence of $O_n(t)$ to $O(t)$ is achieved for small values $n\gtrsim 1$.
In panel (b), we represent the same quantities for a distant operator $j=7$. In this case, the dynamics of the OTOC includes a long time slow behavior, which we attribute to the diffusive character of the operator $W(t)$~\cite{VonKeyserlingk2018,Nahum2018}.  This additional complexity of the operators [compared to ballistic spreading as in (a)] is captured for spatial resolutions $n\gtrsim 4$. In panel (c), we show another example of deviation from ballistic scrambling, with strong oscillations of the OTOCs which are quantitatively captured for $n\ge 2,3$.
Finally, we show in panel (d) the different regimes of scrambling as a function of the transverse field $h_x$ for a fixed large time $Jt=35.2$. \emph{All modified OTOCs identify a region of fast scrambling around $h_x\sim J$}. Interestingly, the modified OTOCs with $n\ge 1$ provide excellent approximation to $O(t)$ also in a regime of slow scrambling $h_x \ge 1.25J$. Finally, for $h_x \le 0.75J$, an increased resolution is necessary to access $O(t)$.
\emph{This example shows that the required resolution $n$ to access $O(t)$ up to a given error depends on the type of evolution realized by $U(t)$, and is generically small.}
Also note that these results also suggest that the convergence properties of the series $O_n(t)$ can be useful to identify different regimes of scrambling.


\subsection{MBL dynamics with modified OTOCs}\label{sec:MBL}
As a second example, with opposite type of scrambling, we consider MBL, which is the paradigmatic example of a closed quantum system which does not
thermalize~\cite{Basko2006,Nandkishore2015}. As a key signature, the decay of $O(t)$
is \emph{slow}: it occurs at a characteristic time $t_B$ which scales exponentially
with the distance $r$ between the support of $W$ and
$V$ at $t=0$~\cite{Fan2016,Chen2017}, which has to be contrasted to the linear increasing
$t_B\propto r $ of chaotic systems.

\begin{figure}
  \includegraphics[width=\columnwidth]{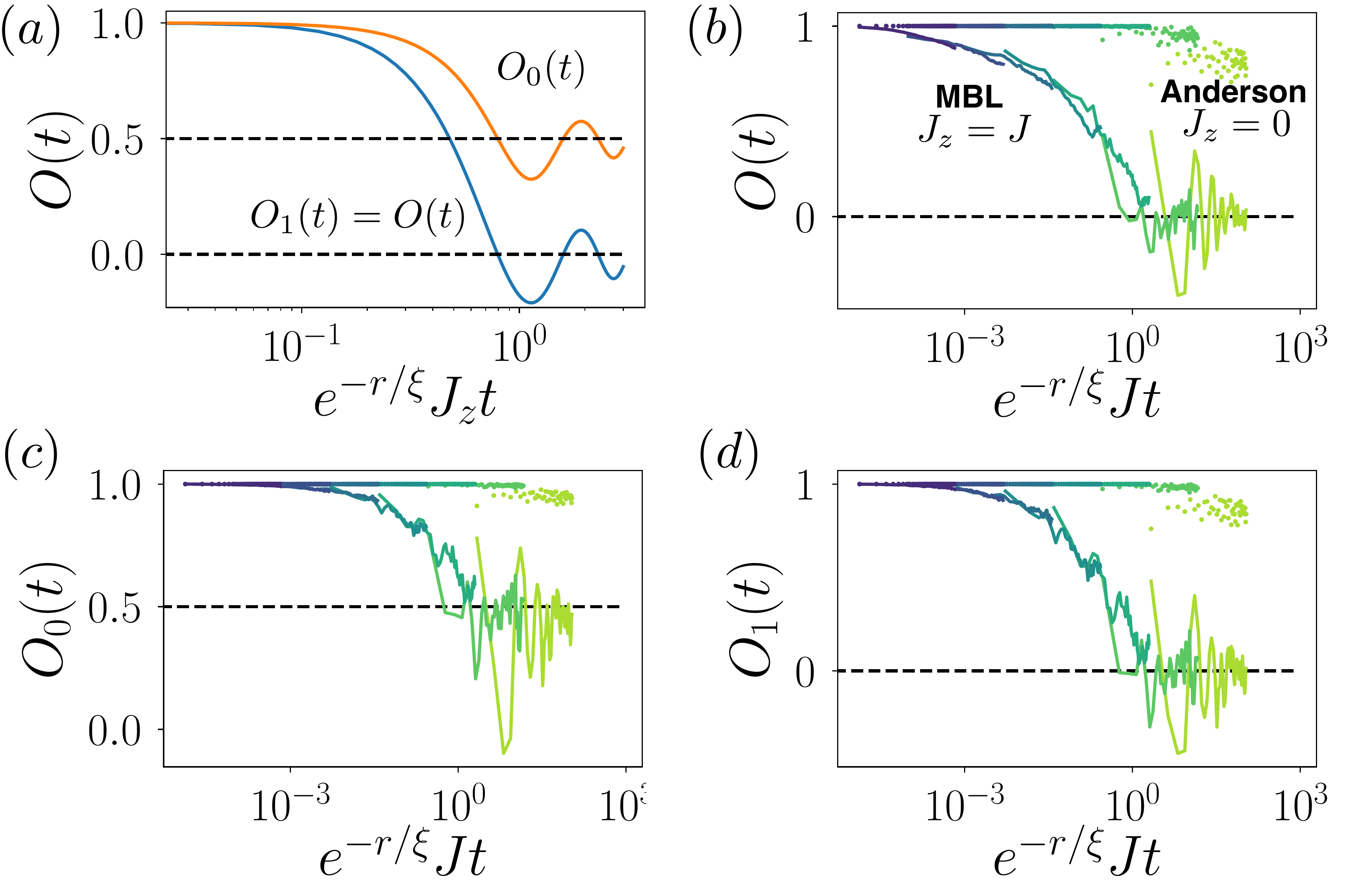}
  \caption{{\it MBL dynamics.}
    (a) Values of $O(t)$ and $O_{0,1}(t)$ in the $\ell$-bit model.
    (b-d) Numerical simulations for a disordered XXZ chain showing that the generic
     and scaling of $O_0(t)$ match $O(t)$, while $O(t)\approx O_1(t)$. We used $20$ random realizations of the disorder.  Green to purple curves represent increasing values of $j=2,\dots,N$. The lines
    (circles) correspond to MBL $J_z=J$ (Anderson
    $J_z=0$) dynamics, respectively. The $x$-axis was rescaled using a fitted localization length
    $\xi=2$. Other parameters: $N=8$, $\Delta=10J$. The dashed lines represent the predictions $O(\infty)=0$ and $O_0(\infty)=1/2$.
    \label{fig:MBL}}
\end{figure}

To begin our analysis, we first calculate the modified OTOC
$O_{0,1}(t)$ for the phenomenological
$\ell$-bit model~\cite{Serbyn2013,Huse2014} described by the Hamiltonian
\begin{equation}
H = \sum_i h_i^z \sigma_i^z + \sum_{i<j} J_{\mathrm{R},ij} e^{-|j-i|/\xi} \sigma_i^z \sigma_j^z, 
\end{equation}
with $h_i^z$ random fields, $J_{\mathrm{R},ij}$ random interactions strengths which are taken uniformly in $[-J_z,J_z$], and $\xi$ the localization length. Here, we only consider $2$-body interaction terms, which is sufficient to show that MBL exhibits slow scrambling~\cite{Fan2016,Chen2017}.
With $U(t)=e^{-iHt}$, $W=\sigma_j^x$, and
$V=\sigma_1^x$, one finds~\cite{Fan2016}
\begin{eqnarray}
    O(t)&=&  \mathrm{sinc}\left(4 J_z e^{-r/\xi}t\right) \nonumber \\
    O_\mathrm{0}(t) &=& \frac{1+2 O(t)}{2+O(t)}, \quad O_1(t)=O(t), \label{eq:lbits}
\end{eqnarray}
 with $r=j-1$. Thus, for slow MBL scrambling, \emph{$O_0(t)$ is related to $O(t)$ via a simple transformation, while we obtain an exact equivalence between the one-site resolved modified OTOCs $O_1(t)$ and $O(t)$}.
In the non-interacting case $J_z=0$, each $\ell$-bit evolves independently $O(t)=O_\mathrm{0,1}(t)=1 = \mathrm{const.}$ 
With interactions $J_z>0$, the decay of both $O(t)=O_1(t)$, and $O_0(t)$ occurs at $t_c=e^{r/\xi}/J_z$. This  analytical result is shown in Fig.~\ref{fig:MBL}(a).
Note that similar to the case of chaotic dynamics, $O(t)$ tends to zero at long times while $O_0(t)$
saturates to a finite value, here $1/2$. These results are confirmed by our numerical
simulations of a disordered XXZ chain~\cite{Pal2010} shown in Fig.~\ref{fig:MBL}(b-d). For the non-interacting case $J_z=0$, both OTOCs values remain close to $1$. In the MBL phase, the decay of $O_\mathrm{0,1}(t)$ exhibits the expected exponential scaling with the distance $r$, and converges at long times to the predicted values of $1/2$, $0$, respectively. \emph{As predicted by the $\ell$-bit model, the values of $O(t)$ and $O_1(t)$ are almost identical.}

\subsection{Information scrambling by long-range interactions with modified OTOCs}\label{sec:Longrange}
So far we have considered examples where interactions were local, with analytical models supporting the statement that low-order $n$ modified OTOCs provide good approximations of $O(t)$. 
To conclude, we study numerically scrambling in a long range interacting model. This situation is particularly interesting as the decay of the OTOCs is not necessarily controlled by a butterfly velocity associated with the presence of a light cone~\cite{Tran2018,Luitz2018}.

Here, we consider a long-range XY model, realized for instance in trapped ion experiments~\cite{Blatt2012}, with time evolution operator
 $U(t)=\exp(-i H_\mathrm{LR} t)$,  and 
\begin{equation}
H_\mathrm{LR} = \sum_{j>i} \frac{J}{(j-i)^\alpha} \left( \sigma_i^+ \sigma_j^- + \mathrm{h.c}\right), 
\end{equation}
with $\alpha$ a positive number controlling the range of the interactions. 

\begin{figure}
\includegraphics[width=\columnwidth]{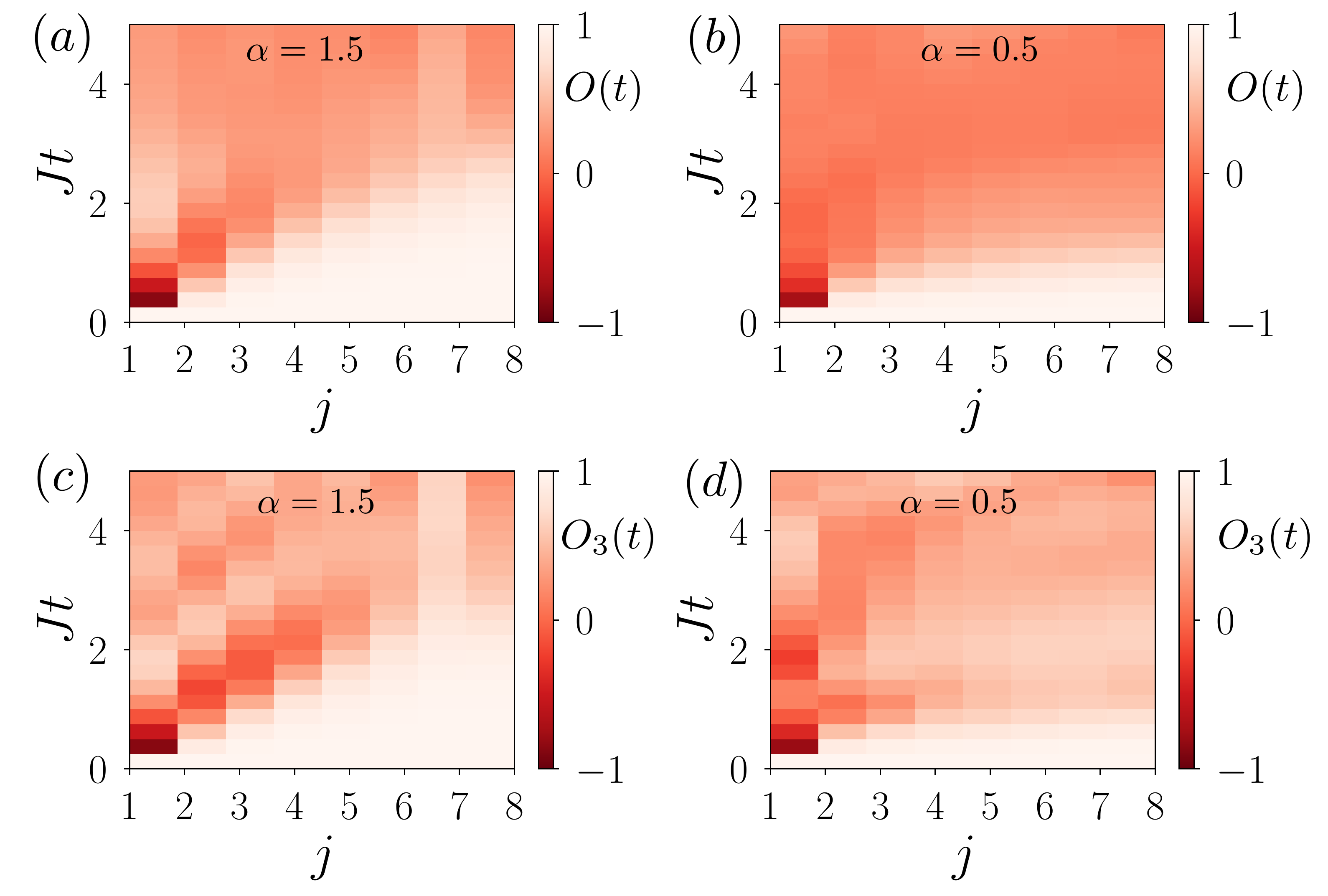}
\caption{{\it Scrambling with long-range interactions}\label{fig:Longrange}
(a-b) OTOC $O(t)$ for $N=8$, and two values of  $\alpha=1.5,0.5$.
(c-d) Same as (a-b) for the modified OTOC $O_3(t)$, which also distinguishes the two regimes of scrambling.
}
\end{figure}

We show in Fig.~\ref{fig:Longrange} the space-time expansion of $O(t)$ and $O_3(t)$ for two different values of $\alpha$. For $\alpha=1.5$, the system satisfies Lieb-Robinson bounds~\cite{Tran2018,Luitz2018}  meaning that the system behaves effectively as if the interactions were local. This manifests by a light-cone spreading of the OTOCs. Conversely, for $\alpha=0.5$, the characteristic decay time of $O(t)$ is superlinear with respect to the operator position $j$. These two types of behaviors are well captured by the modified OTOC $O_3(t)$. 

\section{Errors and imperfections}\label{sec:errors}
We conclude our manuscript by presenting a study of errors and imperfections. From this analysis, we can draw the conclusion that OTOCs can be measured with good precision in AMO and superconducting qubit experiments with current technology, and with total number of measurements compatible with state-of-the-art repetitions rates~\cite{Brydges2018}.

\subsection{Statistical errors}Statistical errors arise in an experiment from a finite number $N_M$ of measurements per random unitary $u$ to access the expectation values $\langle W(t) \rangle_{u,k_0}$ and $\langle V^\dagger W(t) V \rangle_{u,k_0}$, and from a finite number of random unitaries $N_u$ used to estimate the correlation coefficients. This results in deviations $\mathcal{E}=|[O(t)]_e -O(t)|$ between estimated and exact correlation coefficients.

\begin{figure}
\includegraphics[width=\columnwidth]{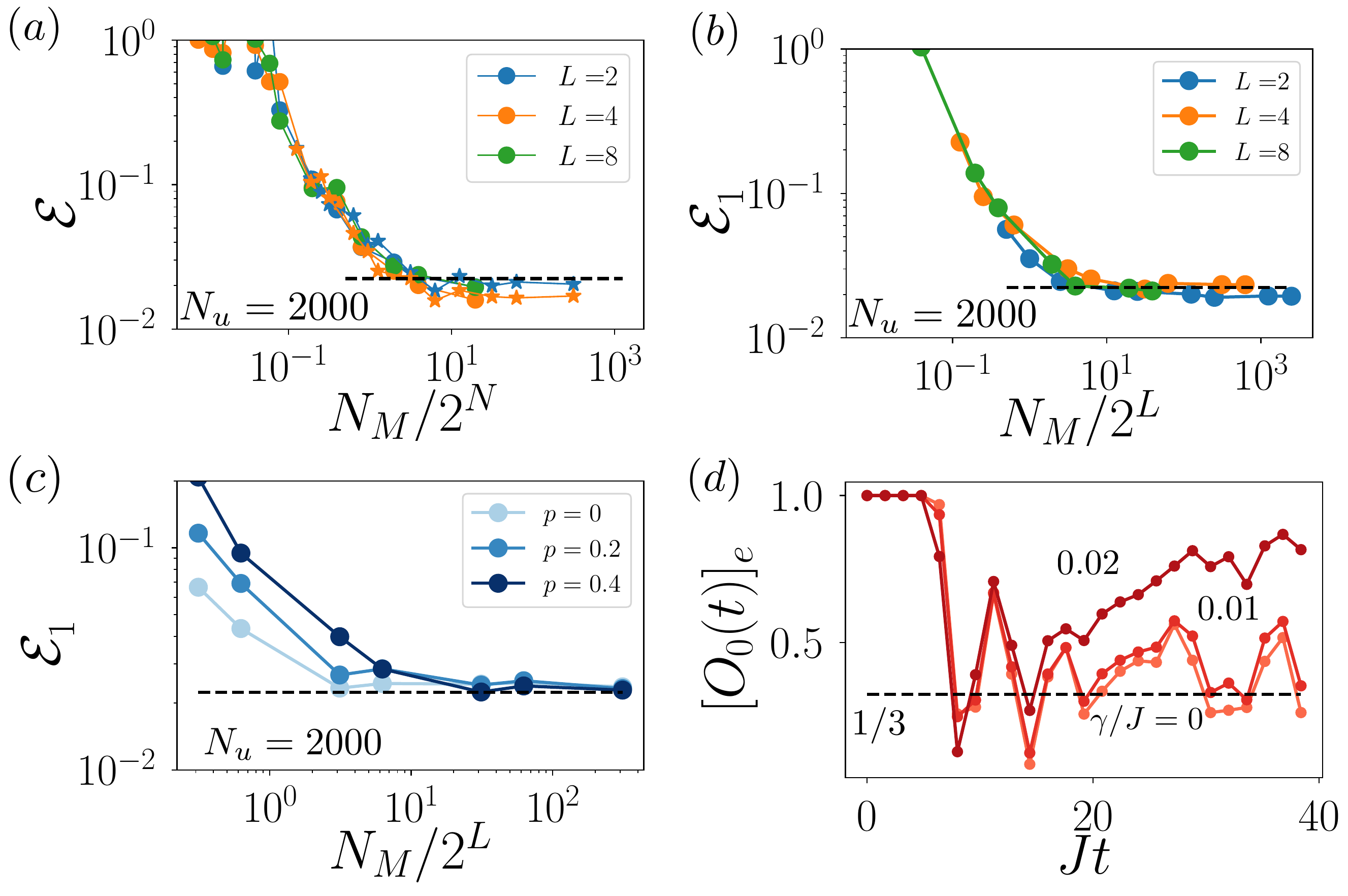}
\caption{{\it Statistical errors, imperfections and decoherence.}
   (a-b) Statistical errors for global and local protocols in the case of the scrambling model Eqs.~\eqref{eq:scrambling1},~\eqref{eq:scrambling2}. For $N_M\ll 2^N$ (a), resp. $N_M\ll 2^L$ (b), the errors reach a statistical plateau at $1/\sqrt{N_u}$. For the simulations of the global protocols, with use two values of $N=4,8$ (circles, stars).
   (c) Effects of depolarization on statistical errors: this decoherence mechanism rescales the values of the observables, this implies that one should perform a slightly larger number of measurements to access $O_1(t)$ with a given precision.
   (d) Decoherence vs.\ scrambling in the kicked Ising model. Each qubit is subject to
  spontaneous emission with rate $\gamma$.  Parameters are $N=6$, $j=4$, $h_x=J$, $h_z=0.809J$, $N_M=\infty$, $N_u=100$.
  The effect of decoherence manifests by an increase of the correlations at long times.
  \label{fig:statistics}}
\end{figure}

The scaling of statistical errors can be explained best in terms of the
phenomenological model for scrambling Eqs.~\eqref{eq:scrambling1},\eqref{eq:scrambling2}. Accordingly, using
global unitaries, the typical value of
\mbox{$\langle {W(t)} \rangle_{u,k_0} \sim 1/\sqrt{2^{N}}$} is suppressed
exponentially. 
The number of projective measurements $N_M$ required
to access $O(t)$ up to a given error thus scales as
$2^{N}$.  The protocol based on local unitaries accessing $O_n(t)$ has a crucial advantage, provided low resolution $n\ll N$ is sufficient to approximate $O(t)$. The typical value of the expectation values
$\langle {W(t)} \rangle_{u,k_s} \sim 1/\sqrt{2^{L(t)}}$ scales instead with the
effective complexity $2^{L(t)}$ of the operator. \emph{Accordingly, the early-time dynamics of large chaotic
systems subject to Lieb-Robinson bounds~\cite{VonKeyserlingk2018,Nahum2018}, but also the long-time evolution of an MBL system, both of which
correspond to small scrambling lengths $L(t)\ll N$, are accessible with a
moderate number of measurements $\sim 2^{L(t)}$}. These findings are confirmed by the numerical
simulations of Fig.~\ref{fig:statistics} showing the statistical error $\mathcal{E}$ for the global protocol [panel (a)] and the corresponding error $\mathcal{E}_{1}$  for the local protocol with $n=1$ [panel (b)]. Note that for both protocols, when convergence with respect to $N_M$ is reached, the typical statistical error is $1/\sqrt{N_u}$ (consistent with the central limit theorem).

\subsection{Imperfections and decoherence}
Our protocol has a certain robustness against various types of experimental imperfections and decoherence.
For instance, the detrimental effect of the imperfect implementation of local
random unitaries $u$ is strongly suppressed when performing the ensemble
average, c.f. App.~\ref{app:errors}.
The protocol is also robust against readout errors.
In the case of decoherence, depolarization noise only
rescales the values of the measurements of $W$, while leaving statistical
correlations unaltered, see App.~\ref{app:depo} and Fig.~\ref{fig:statistics}(c).
\emph{For other sources of decoherence, the values of correlations can be affected, but in a way
which can be clearly distinguished from unitary scrambling}. While scrambling
generically leads to a \emph{decay} of statistical correlations, decoherence
\emph{increases} correlations. This behavior, which is opposite to the case of protocols based on the reversal of
time evolution~\cite{Swingle2018,Yoshida2018,Landsman2018}, can be understood by noting that, for a
Markovian dissipative evolution, the distance between two different states
always decreases with time~\cite{Breuer2009}. 
This is illustrated in
Fig.~\ref{fig:statistics}(d) for the estimation of the modified OTOC $[O_0(t)]_e$.

\section{Conclusion and outlook}
In the present work we provide novel protocols to measure OTOCs for spin
models, based on statistical correlations between measurement outcomes obtained
from random initial states from both {\em global} and {\em local} random unitaries. These protocols can be implemented in
state-of-the-art quantum simulation experiments on various physical platforms,
in particular in Rydberg atoms, trapped ions, or superconducting qubits, which
provide high repetition rates.  The paradigm of extracting non-standard
correlation functions of quantum many-body systems from statistical correlations
points to several interesting future developments. This includes extensions to
measure modified OTOCs for Hubbard models~\cite{Bloch2012}.
We also present
indications in App.~\ref{app:thermal} that the protocol can be adapted to access OTOCs for thermal
states. This could be used to extract the crucial temperature dependence of the
Lyapunov exponents in models of high-energy
physics such as SYK~\cite{Sachdev1993, Kitaev, Hayden2007, Shenker2014, Maldacena2016,
  Banerjee2017}.

\begin{acknowledgements}We thank M.~Heyl, E.~Altman, M.~Dalmonte, T.~Schuster, C.~Roos, C.~Maier, T.~Brydges, and M.~Joshi for discussions and comments on the manuscript. Work in Innsbruck is supported by the ERC Synergy Grant UQUAM and the SFB FoQuS (FWF Project No. F4016-N23).  NYY acknowledge support from the DOE under contract PHCOMPHEP-KA24 and the Office of Advanced Scientific Computing Research, Quantum Algorithm Teams Program. Numerical simulations were realized with QuTiP~\cite{ Johansson20131234}.
\end{acknowledgements}
\appendix
\section{Correspondence between statistical correlations and OTOCs with local unitaries}\label{app:local}
\begin{figure}[t]
\includegraphics[width=\columnwidth]{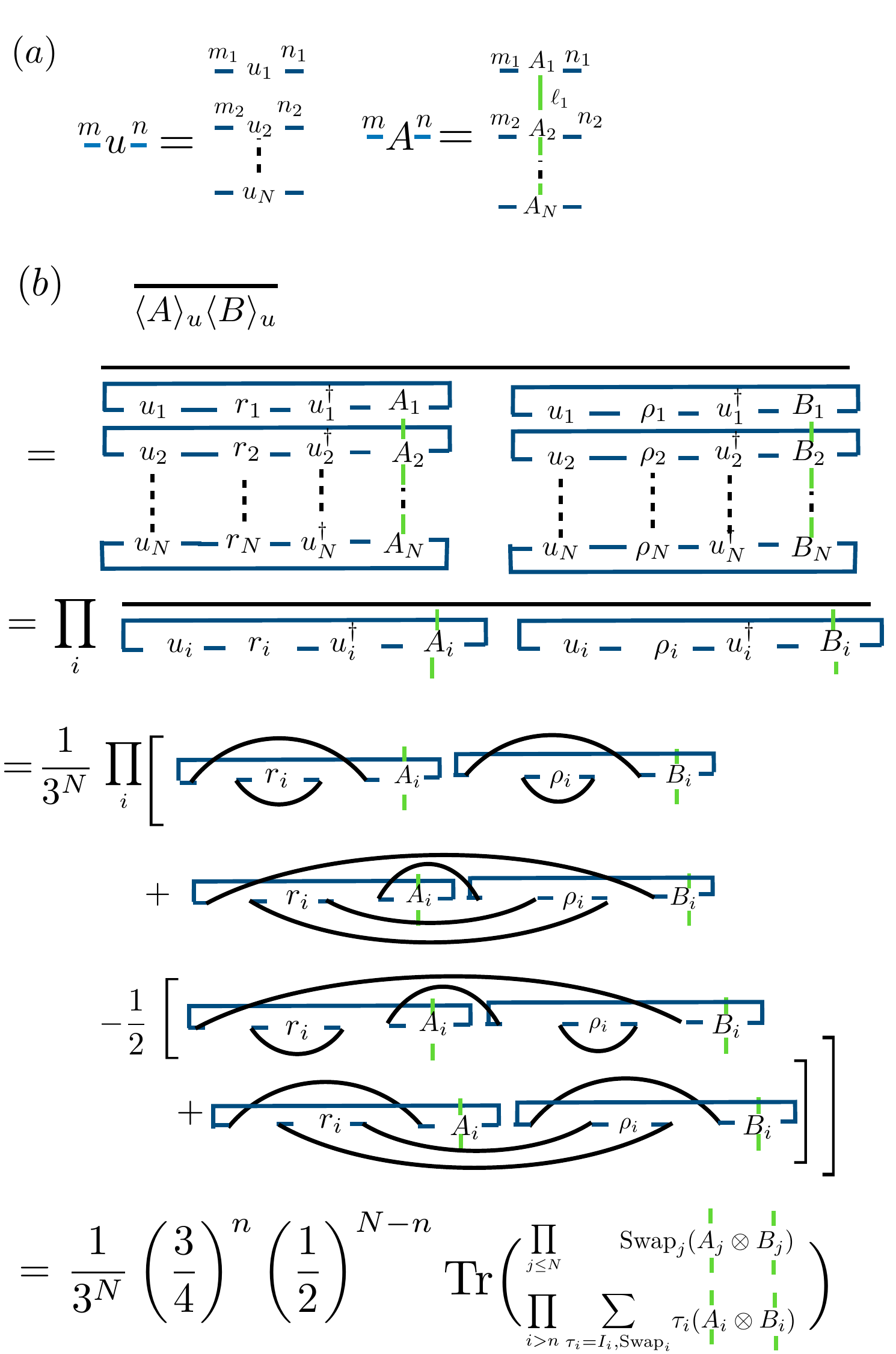}
\caption{{\it Identities for our protocol with local unitaries.} (a) MPO representation of multi-site indexed operators. (b)  Correlations between two measurements with $A=W(t)$, $B=V^\dagger W(t) V$.\label{fig:Proof_L}}
\end{figure}

In this section, we prove the relation between statistical correlations and OTOCs, for the second protocol with local random unitaries  
$u=u_1\otimes \cdots \otimes u_N$, where each $u_i$ is sampled independently from $\mathrm{CUE}(d)$, $d$
being the local Hilbert space dimension ($d=2$ for the spins $1/2$ considered here). 
Here, we extend our diagrammatic approach introducing a Matrix-Product-Operator
(MPO)~\cite{Schollwock2011} representation for the operators $W(t)$ and $V$.
This is shown in Fig.~\ref{fig:Proof_L}: each (blue) physical index is
contracted following the 2-design rule shown in panel (a) while the (green) bond
links remain unchanged. To simplify the derivation of the proof, we rewrite $O_n(t)$ as 
\begin{eqnarray}
O_n(t) \equiv \frac{\overline{\langle W(t) \rangle_{u,n} \langle V^\dag W(t)V \rangle_{u,k_0}}}{\overline{\langle W(t) \rangle_{u,n} \langle W(t) \rangle_{u,k_0}}}, 
\end{eqnarray}
with $\langle W(t) \rangle_{u,n}\equiv \sum_{k_s\in E_n } (-1/2)^{d[k_{s},k_0]}  \langle W(t) \rangle_{u,k_{s}}=\mathrm{Tr}(r_n W(t))$, with the operator $r_{n}=r_{1,n}\otimes \dots\otimes r_{1,N}$ written as a tensor product with $r_{i,n}=\ket{k_0^{(i)}}\bra{k_0^{(i)}}-1/2 \left(\sigma_i^x\ket{k_0^{(i)}}\bra{k_0^{(i)}}\sigma_i^x\right)\delta_{i\le n}$. The operator $r_n$ gathers all the information about the initial states $\ket{k_s}\in E_n$, and corresponding chosen weights $c_{k_s}=(-1/2)^{d[k_0,k_s]}$.
Finally, we also use the tensor product decomposition $\rho_0=\rho_1\otimes\dots\otimes\rho_N$, with $\rho_i = \ket{k_0^{(i)}}\bra{k_0^{(i)}}$. 

 We can now prove graphically in Fig.~\ref{fig:Proof_L}
 \begin{eqnarray}
 &&\overline{\langle W(t) \rangle_{u,n} \langle VW(t)V \rangle_{u,k_0}} 
 \\
  &=& \frac{1}{3^N}\left(\frac{3}{4}\right)^n\left(\frac{1}{2}\right)^{N-n}\sum_{\tau\in \mathcal{E}_n^\mathrm{(L)}} \mathrm{Tr} (\tau W(t) \otimes VW(t)V),
   \nonumber
 \end{eqnarray}
 where the index $n$ is omitted in the graphics. 
The ensemble $\mathcal{E}_n^\mathrm{(L)}$ consists of all the $2^{N-n}$ permutations of the form $\tau=\prod_{j\ge n} \mathrm{Swap}_j \prod_{i>n}  \tau_i$, $\tau_i 
=I_i,\mathrm{Swap}_i$, with the local swap operator \mbox{$\mathrm{Swap}_i\ket{k_i}\otimes\ket{k'_i}=\ket{k'_{i}}\otimes\ket{k_{i}}$}. This leads directly to the desired equality
\begin{eqnarray}
O_n(t)&=& \frac{ \sum_{\tau \in \mathcal{E}_n^{(\mathrm{L})}}
\mathrm{Tr}[\tau (W(t)\otimes V^\dagger W(t) V)]}{\sum_{\tau \in \mathcal{E}_n^{(\mathrm{L})}}
\mathrm{Tr}[\tau (W(t)\otimes W(t))]}
\nonumber \\
&=&\frac{\sum_{A, \{1\dots n\} \subseteq A} \tr_{A} \!\left( W_A(t) [V^\dagger W(t) V]_A  \right)}{\sum_{A, \{1\dots n\} \subseteq A} \tr_A\! \left( W^2_A(t) \right)}.\label{eq:identityOL_app}
\end{eqnarray}
In particular for $n=N$, we obtain $O_N(t)=O(t)$.

\section{OTOCs for Haar scrambling}\label{app:haar}
\begin{figure}
\includegraphics[width=0.9\columnwidth]{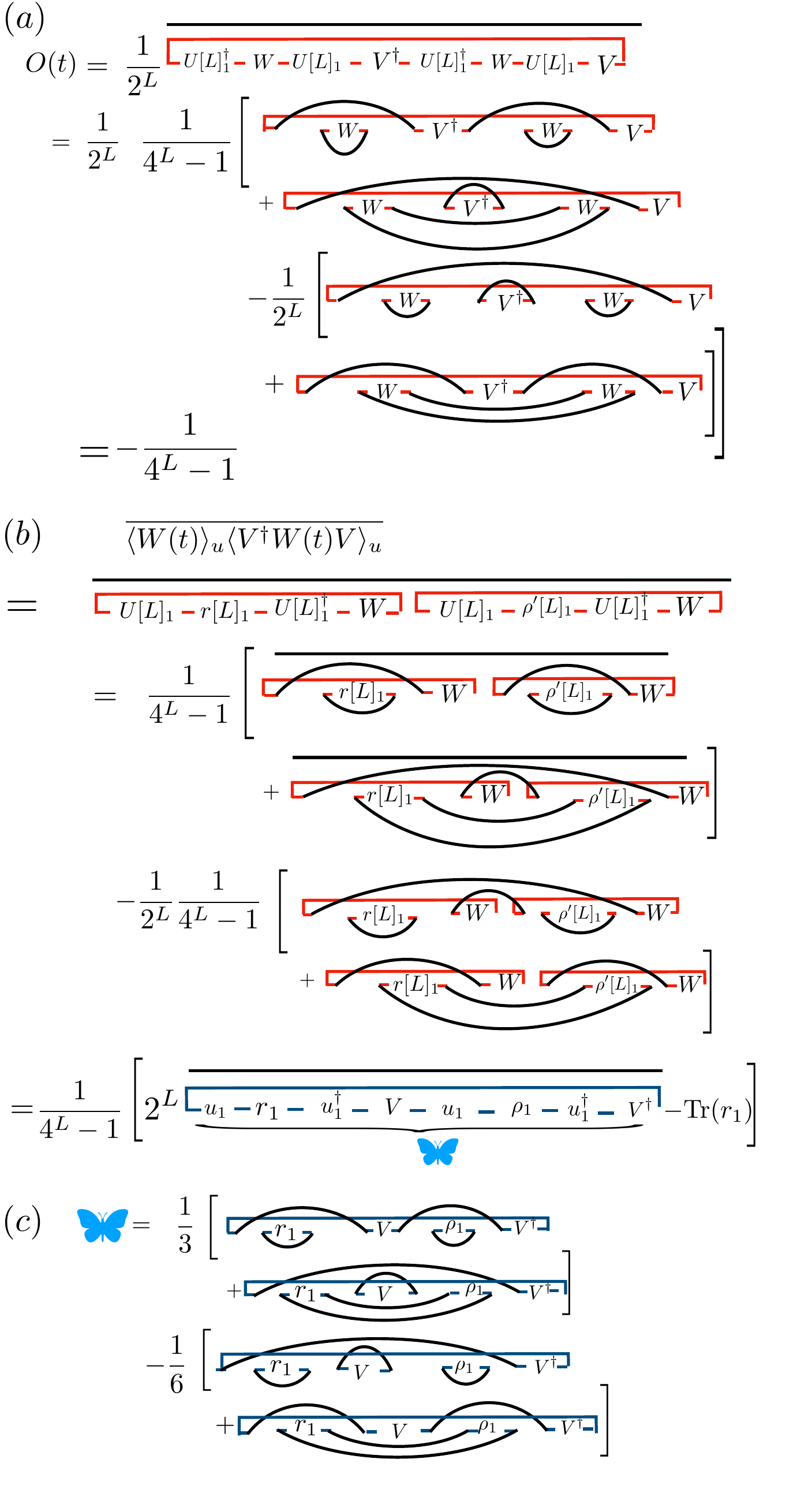}
\caption{{\it Proof of Eq.~\eqref{eq:scrambling1} using a diagrammatic approach.}
In panels (a-b), the ensemble average is performed over the unitaries $U$ leading to the contractions of the red indices.
In panel (c), the ensemble average is taken with respect to the unitaries $u_1$ with the contracted indices shown in blue.  \label{fig:proofhaar}}
\end{figure}

We now prove Eqs.~\eqref{eq:scrambling1},~\eqref{eq:scrambling2} of the main text. The case $j>L(t)$ is straightforward due
to the commutativity of $V$ and $W$. For $j\le L(t)$, the equality written for
the OTOC $O(t)$ follows directly from the $2$-design identities
[c.f. Fig.~\ref{fig:proofhaar}(a)], and can also be found in
Ref.~\cite{Roberts2016}. 

For the modified OTOCs, we use the mapping to statistical correlations to calculate the values corresponding to Haar scrambling. We show in Fig.~\ref{fig:proofhaar}(b-c) the
evaluation of
$\overline{\langle W(t) \rangle_{u,n=0,1} \langle V^\dagger W(t) V \rangle_{u,k_0}}$, which
can be adapted to derive $\overline{\langle W(t) \rangle_{u,n}\langle W(t) \rangle_{u,k_0}}$ by replacing $V$
by the identity matrix. 
Since $u$ is a product of local random unitaries, we could use the decomposition $X = X[L]_1\otimes \dots$, 
$X[L]_1 = \otimes_{i\le L} u_i X_i u_i^\dag$, for $X=r,\rho$. We also used the notation $\rho'[L]_1 =V\rho[L]_1 V^\dag$.
This proves the desired identities
\begin{eqnarray}
O_0(t)&=&\frac{2^L/3-1}{2^L-1}
\quad
O_1(t)=\frac{-1}{2^{L+1}-1}.
\end{eqnarray}

\section{OTOCs for many body localization}
To describe analytically the behavior of OTOCs in the MBL
phase~\cite{Fan2016,Chen2017}, we consider the $\ell$-bit model with time
evolution $U(t)=\exp(-iHt)$ described by the Hamiltonian
\begin{equation}
H = \sum_i h_i^z \sigma_i^z + \sum_{i<j} J_{i,j}  \sigma_i^z \sigma_{j}^z, 
\end{equation}
where $h_i^z$ are random fields which as we show below have no effect on the
OTOC, $J_{ij}=J_{\mathrm{R},ij} \exp(-|j-i|/\xi)$, $ J_{\mathrm{R},ij}$ a random
interaction amplitude with probability distribution $f(J_{\mathrm{R},ij})$ which
we assume uniform in $[-J_z,J_z]$, and $\xi$ the localization length. We study
the behavior of the OTOCs for the operators $W=\sigma_j^x$, and
$V=\sigma_1^x$. In the Anderson case $J_z=0$, each $\ell$-bit evolves independently,
and we have $O(t)=O_n(t)=1$.

\subsection{Single disorder realization}
We now address the general MBL case $J_z>0$ and first consider a single disorder realization of the interaction matrix $(J_{ij})$, writing  the Heisenberg operators as
\begin{eqnarray}
W(t) &=& \sigma_j^x  e^{-2it h_j^z \sigma_j^z} e^{-2it \sum_{i\neq j} J_{ij} \sigma_i^z \sigma_j^z} \nonumber \\
V^\dagger W(t) V &=& \sigma_j^x   e^{-2it h_j^z \sigma_j^z}e^{-2it \sum_{i \neq j} J_{ij} c_i \sigma_i^z \sigma_j^z}, 
\end{eqnarray}
with $c_{i\neq 1}=1$, $c_1=-1$. We then obtain
\begin{eqnarray}
O(t) &=& \frac{1}{2^N} \mathrm{Tr}(\exp(-4it J_{1j}\sigma_1^z \sigma_j^z)) = \cos(4 J_{1j}t), 
\end{eqnarray}
as already shown in Ref.~\cite{Fan2016,Chen2017}. We now evaluate $O_{n=0,1}(t)$ using the first line given  in Eq.~\eqref{eq:identityOL_app}. We first perform the trace operation with respect to the site $j$
\begin{eqnarray}
&&\mathrm{Tr}_j \left[ (I_j+\mathrm{Swap}_j) (W(t) \otimes W(t))\right]\nonumber \\
 &&=2\cos \left(2t \sum_{i\neq j} J_{ij} (\sigma^z_i- \tilde\sigma_i^z)\right)\nonumber \\
&&\mathrm{Tr}_j \left[(I_j+\mathrm{Swap}_j) (W(t) \otimes V^\dagger W(t) V)\right]\nonumber \\
&&= 2\cos \left(2t \sum_{i\neq j} J_{ij}  (\sigma^z_i- c_i \tilde\sigma_i^z)\right),
\end{eqnarray}
with $\tilde\sigma_i^\beta$ ($\beta=x,y,z$) the set of Pauli matrices in the
``copy'' Hilbert space $\mathcal{H}_i$ associated to site $i$. We can then
calculate the trace over the remaining sites, for instance:
\begin{eqnarray}
&&\mathrm{Tr}_k \left[(I_k+\mathrm{Swap}_k ) \cos\left(2t \sum_{i\neq j} J_{ij} (\sigma^z_i- \tilde\sigma_i^z)\right) \right]\nonumber\\
&&=\left[ 4\cos(2t J_{kj})^2+2 \right] \cos\left(2t \sum_{i\neq {k,j}} J_{ij} (\sigma^z_i- \tilde\sigma_i^z)\right)
\\
&&\mathrm{Tr}_k \left[(I_k+\mathrm{Swap}_k) \cos\left(2t \sum_{i\neq j} J_{ij} (\sigma^z_i-c_i \tilde\sigma_i^z)\right)  \right] \nonumber\\
&&= \left[ 4\cos(2t J_{kj})^2+2\cos(2(1-c_k)J_{kj}t) \right]
\nonumber \\
&&\cos\left(2t \sum_{i\neq k,j} J_{ij} (\sigma^z_i- \tilde\sigma_i^z)\right).
\end{eqnarray}
All the factors which enter in the numerator and in the denominator in Eq.~\eqref{eq:identityOL_app} are identical, except for the position $k=1$ of the $V$ operator. This leads to 
\begin{eqnarray}
O_{0}(t) &=& \frac{ 4\cos(2 J_{1j}t)^2+2\cos(4J_{1j}t)}{4\cos(2 J_{1j}t)^2+2}
= \frac{ 2\cos(4 J_{1j}t)+1}{\cos(4 J_{1j}t)+2}.
\nonumber \\
O_{1}(t) &=& \cos(4 J_{1j}t).
\end{eqnarray}

\subsection{Averaged OTOCs}
Considering now a distribution of the random realizations of $J_{ij}$,  $O_{n=0,1}(t)$ is now obtained from Eq.~\eqref{eq:identityOL_app}, where the numerator and the denominator are replaced by their ensemble average.
Repeating the above derivation, replacing each cosine contribution by the average
\begin{equation}
 \cos(\alpha J_{ij}t )\to \int_{-J_z}^{J_z} dJ_{\mathrm{R},ij} \cos(\alpha J_{\mathrm{R},ij} e^{-|j-i|/\xi} t),
 \end{equation}
we obtain Eq.~\eqref{eq:lbits}.

\subsection{Simulations of the disordered XXZ model}
To describe the OTOCs in the MBL dynamics~\cite{Fan2016}, we consider the XXZ model
\begin{equation}
H = \sum_i \left[J\left(\sigma_i^+ \sigma_{i+1}^- +\sigma_i^- \sigma_{i+1}^+\right) +J_z \sigma_i^z \sigma_{i+1}^zx \right]+  \sum_i h_z^i \sigma_i^z, 
\end{equation}
with $h_z^i$ sampled from a uniform distribution $[-\Delta,\Delta]$.

\section{Numerical simulation with decoherence}

To study the competition between scrambling dynamics and decoherence, we considered the kicked Ising model. The dynamics was calculated from the Lindblad master equation
\begin{eqnarray}
&&\dot \rho(t) = -i [H(t) ,\rho(t)] + \gamma \sum_i\mathcal{L}[\sigma_i^-](\rho(t)) \nonumber \\
&&H(t \in[(n-1)T,nT-\frac{T}{2}] )= h_x \sum_i  \sigma_i^x 
\nonumber \\
&&H(t \in[nT-\frac{T}{2},nT] )=J\sum_i \sigma^z_i \sigma^z_{i+1}+\sum_i h_z  \sigma_i^z 
\nonumber \\
&&\mathcal{L}[\sigma_i^-](\rho) = \frac{1}{2} \left[2\sigma_i^- \rho \sigma_i^+ -\rho \sigma_i^+\sigma_i^- -  \sigma_i^+\sigma_i^- \rho \right],
\end{eqnarray}
with initial condition $\rho(0)=\rho_0$, and $\gamma$ the spontaneous emission decay rate (identical for each qubit).

\section{Numerical study of statistical errors}\label{app:statistics}

\begin{figure}
\includegraphics[width=1\columnwidth]{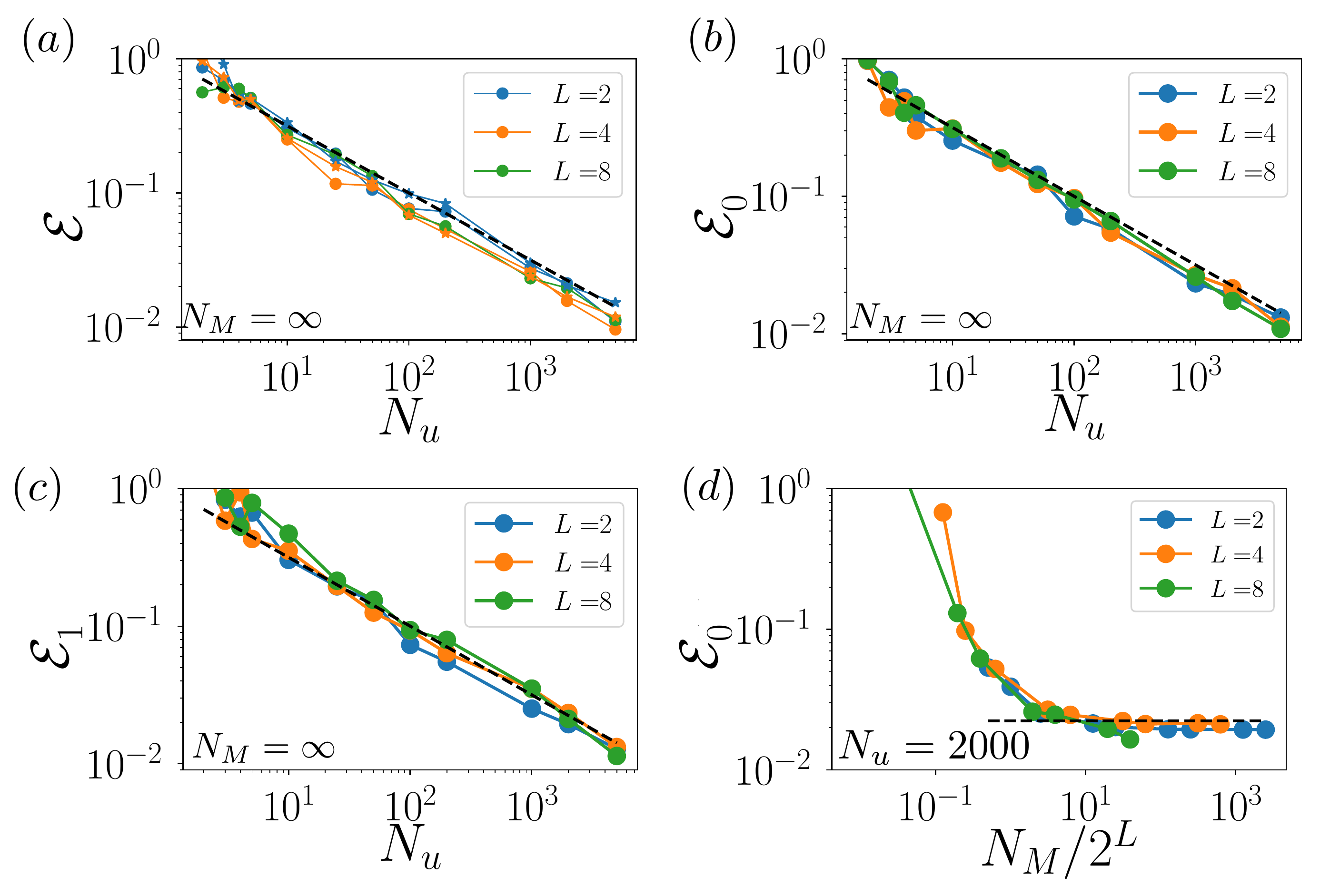}
\caption{{\it Additional simulations on statistical errors}  
(a) Scalings of the error $\mathcal{E}$ in the estimation of $O(t)$ as a function of $N_u$ with global unitaries, and $N_M\to \infty$.
 We used two values of $N=4,8$ (circles, stars). 
  (b-c) Same as (a) for  local unitaries estimating $O_{0,1}(t)$  (here the simulations are independent of $N$). 
  (d) Statistical errors for the estimation of $O_0(t)$ as a function of  $N_M$, and $N_u=2000$. 
 In all panels, the dashed lines represent $1/\sqrt{N_u}$.
  \label{fig:statistics_SM}}
\end{figure}

Here, we present complementary data to Fig.~\ref{fig:statistics}, showing the scaling of statistical errors in our protocol with global and local unitaries. 
The results are shown in Fig.~\ref{fig:statistics_SM}. The data confirm that statistical errors for $N_M\to \infty$ decrease as $1/\sqrt{N_u}$ with growing number of applied random unitaries $N_u$, independently of $L(t)$ and $N$ [panels (a-c)]. Panel (d) shows the scaling of statistical errors with respect to $N_M$ for the first modified OTOC. 

\section{Robustness of the protocols}
In this section we give three example showing how random measurements are robust against different kind of perturbations. We consider in each case the protocol with local unitaries. 

\subsection{Limited reproducibility of generated random unitaries}\label{app:errors}
First we analyze the robustness of our protocol with respect to imperfections in the generation of random unitaries. Specifically, we assume that $\langle W(t) \rangle_{u,k_s}$ is obtained from a random unitary matrix $u=u_1\otimes\cdots u_N$, with $u_i \in \mathrm{CUE(2)}$, while the second measurement $\langle V^\dagger W(t) V \rangle_{u',k_0}$ is obtained from a slightly different unitary $u'=u'_1\otimes\cdots u'_N$ which we write as
\begin{equation}
u_i'= R_{iz}(\theta_{i1})R_{iy}(\theta_{i2})R_{iz}(\theta_{i3})u_i.
\end{equation}
Here, $R_{i\gamma}$ denote single qubit rotations for spin $i$ along the
$\gamma$ axis, and $\theta_{i n}$ are assumed to be random angles drawn
uniformly in $[-\theta,\theta]$, which represent the unwanted mismatch between
the unitaries $u$ and $u'$.

\subsubsection{Analytical understanding of the robustness of the protocol}
To show that our protocol is robust against such imperfections, we calculate the estimator of the OTOC, obtained from 
\begin{equation}
\tilde O_n(t) = \frac{\sum_{k_s\in E_n} c_{k_s} \overline{\langle W(t) \rangle_{u,k_s}\langle V^\dagger W(t) V \rangle_{u',k_0}}}{\sum_{k_s\in E_n}c_{k_s} \overline{\langle W(t)\rangle_{u,k_s}\langle W(t) \rangle_{u,k_0}}}.
\end{equation}
In first order in $\theta_{in}$,  we can write $
u'=u+\sum_{i,n} \theta_{in} A_{in}$, 
with $A_{in}=R'_{in}(0)u$.
This leads to 
\begin{eqnarray}
&&\overline{\langle W(t) \rangle_{u,k_s}\langle V^\dagger W(t) V \rangle_{u',k_0}} = \overline{\langle W(t) \rangle_{u,k_s}\langle V^\dagger W(t) V \rangle_{u,k_0}}
\nonumber \\
&+&\sum_{i,n} \overline{\theta_{in}} \ \overline{\langle W(t) \rangle_{u,k_s}
\mathrm{Tr}([u\rho_0 A_{in}^\dag+A_{in} \rho_0 u^\dag]V^\dagger W(t) V)},\nonumber
\end{eqnarray}
where the second term vanishes due to $\overline{\theta_{in}}=0$. This implies that 
$\tilde O_n(t)=O_n(t)$.  Our protocol is thus
robust to first order in $\theta$ against imperfections of the generated
unitaries. As shown below, the quadratic contribution scales linearly with the
characteristic size $L(t)$ of the operator $W(t)$.

\subsubsection{Numerical example}
We now consider a numerical example for the model of scrambling presented in Sec.~\ref{sec:chaos}, and the estimation of the first modified OTOC $n=0$. To assess the robustness of statistical correlations, we consider $j>L(t)$ so that the exact OTOC is $O_0(t)=1$, and imperfect generated unitaries as written above.

\begin{figure}
\includegraphics[width=0.8\columnwidth]{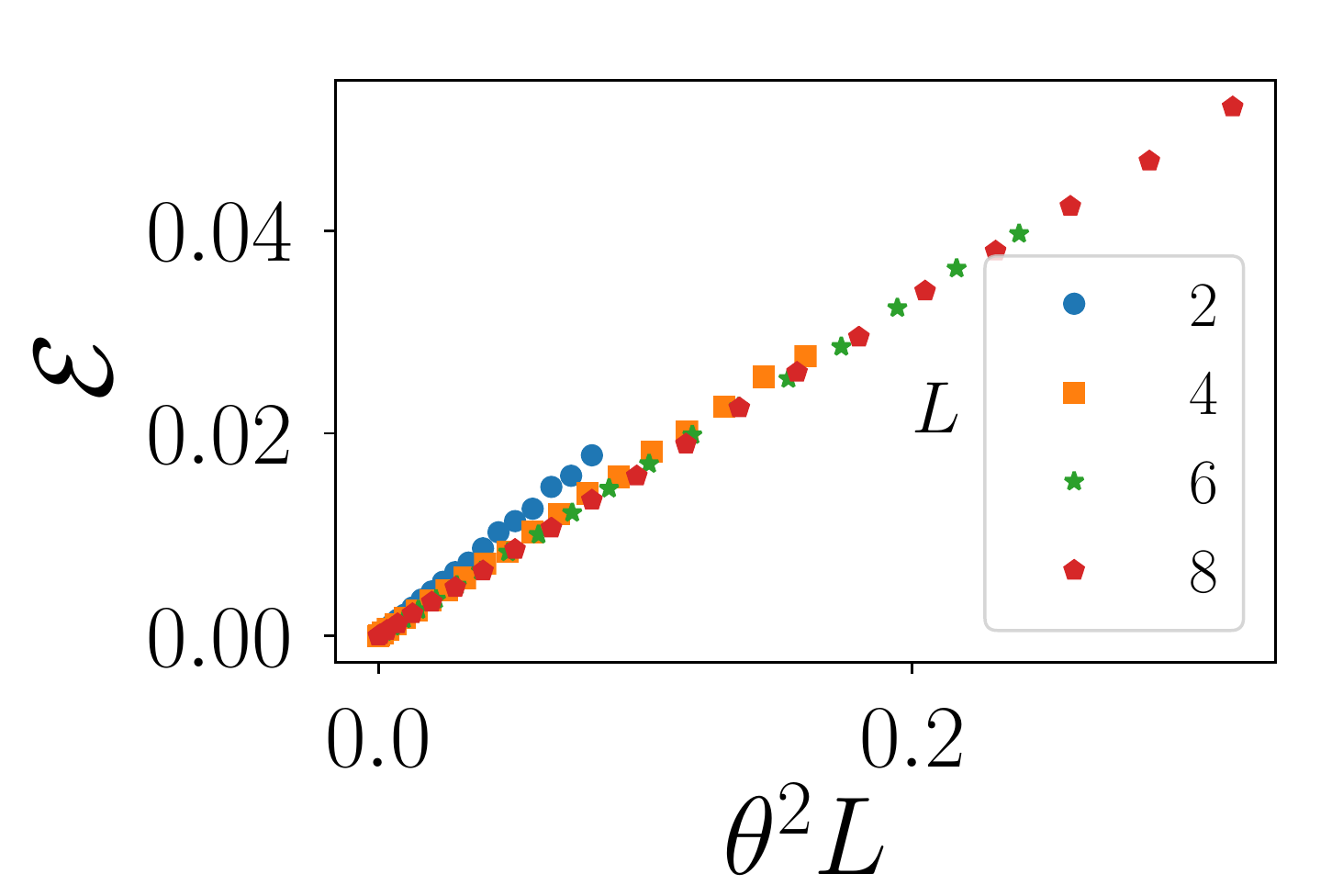}
\caption{Robustness of the protocol with respect to a mismatch between generated random unitaries in the estimation of the first modified OTOC. The error $\mathcal{E}$ is plotted as a function of the scaling parameter $\theta^2 L$, for $L=4,6,8$ and $0<\theta<0.2$ Note that all data points, the error is below $6\%$. \label{fig:Reproducibility}}
\end{figure}

The numerical results are shown in Fig.~\ref{fig:Reproducibility} and confirm our analytical treatment: there is no linear correction in $\theta$ in the error $\mathcal{E}$ of the estimated OTOC. However, the error scales as $\theta^2 L$, showing that the quadratic contribution does not vanish.

\subsection{Robustness against the depolarizing channel}\label{app:depo}
In the presence of depolarizing noise~\cite{Yoshida2018}, the realized quantum state can be written as
\begin{equation}
\rho_f=(1-p) U_\mathrm{tot}(t) \rho_0 U^\dagger_\mathrm{tot}(t) + \frac{pI}{d^N},
\end{equation}
with $0<p<1$ the depolarization probability, $I$ and the identity matrix, and $U_\mathrm{tot}(t)$ the desired time-evolution unitary operator. This translates to the measurement outcomes [for the sequence shown in Fig.~\ref{fig:setup} with $U_\mathrm{tot}(t)=U(t)u$ (a) and $U_\mathrm{tot}(t)=U(t)Vu$ (b)]
\begin{eqnarray}
\langle W(t) \rangle'_{u,k_s} &=& (1-p) \langle W(t) \rangle_{u,k_s} \nonumber \\
\langle V^\dagger W(t) V \rangle'_{u,k_0} &=& (1-p) \langle V^\dagger W(t) V \rangle_{u,k_0}, 
\end{eqnarray}
where we used the fact that $W(t)$ is traceless.
Hence, the two measurements are simply rescaled, and the statistical correlations are not affected by the depolarizing channel. This is illustrated in Fig.~\ref{fig:statistics}(d), where the numerical simulation (for the same model and parameters as in  Fig.~\ref{fig:statistics}(a-c)) shows that the effect of depolarization can be completely eliminated by performing a sufficient number of projective measurements $N_M$.

\subsection{Robustness against readout errors}
We now show that our protocols are also robust against readout errors. Here, we consider for concreteness $W=\sigma_j^z$. 
For each random unitary $u$, and initial state $k_s$ ($k_0$), we assume that each measurement of the operator is estimated  as
\begin{eqnarray}
\langle W(t) \rangle^\mathrm{est}_{u,k_s} &=& 2 P^\mathrm{est}(t,\uparrow,u\ket{k_s})-1 \nonumber \\
\langle VW(t)V \rangle^\mathrm{est}_{u,k_0} &=& 2  P^\mathrm{est}(t,\uparrow,uV\ket{k_0})-1, 
\end{eqnarray}
where $P^\mathrm{est}(t,\uparrow,\ket{k})$ is the estimated probability to detect spin $j$ in state $\uparrow$ after time evolution from the state $\ket{k}$. 
We now consider that  the estimated probabilities $P^\mathrm{est}(t,\uparrow,\ket{k})$ are built from a sequence of $N_M$ measurements
\begin{equation}
P^\mathrm{est}(t,\uparrow,\ket{ k})= \frac{1}{N_M} \sum_i X_i, 
\end{equation}
with $X_i$ the measurement outcomes obtained with error probability $x$:
\begin{eqnarray}
\mathrm{Prob}(X_i=1) &=& (1-x) P(t,\uparrow,\ket{k}) + x  P(t,\downarrow,\ket{k})
\nonumber \\
&=& (1-2x)P(t,\uparrow,\ket{k}) + x.
\end{eqnarray}
In the limit of an infinite number of measurements ($N_M\to \infty$), we obtain $P^\mathrm{est}(t,\uparrow,\ket{ k})=\mathrm{Prob}(X_i=1)$, and thus  
\begin{eqnarray}
\langle W(t) \rangle^\mathrm{est}_{u,k_s} &=& (1-2x) \langle W(t) \rangle_{u,k_s}  \nonumber \\
\langle VW(t)V \rangle^\mathrm{est}_{u,k_0} &=& (1-2x) \langle VW(t)V \rangle^\mathrm{est}_{u,k_0}. 
\end{eqnarray}
Accordingly, as in the case of the depolarizing channel, readout errors simply rescale the value of the observables, i.e., the estimation of $O_n(t)$ is not affected by readout errors.

\section{Thermal OTOCs from global unitaries}\label{app:thermal}

In this section, we show how our protocol can be extended to include  finite temperature corrections to $O(t)$.   To this end, we employ a  high-temperature expansion of the thermal density matrix $\rho_\beta=\exp(-\beta H)/Z$, with $H$ being the many-body Hamiltonian of the system of interest $Z=\textrm{Tr}(\exp(-\beta H))$ and $\beta$ the inverse temperature.
Specifically, instead of considering the canonical finite temperature OTOC 
\begin{equation}
O[\rho_\beta ](t) = \mathrm{Tr}(\rho_\beta W(t) V W(t) V),
\end{equation}
with $\rho_\beta$ a thermal density matrix, we consider a symmetrized variant, introduced in Ref.~\cite{Maldacena2016},
\begin{equation}
O_S[\rho_\beta](t) = \mathrm{Tr}(\rho_\beta^{1/4} W(t)\rho_\beta^{1/4} V\rho_\beta^{1/4} W(t)\rho_\beta^{1/4} V). 
\end{equation}
Here, and in the following, we assume $W(t)$ to be hermitian, $V$ to be
hermitian and unitary, and all operators $W$, $V$, $H$ to be traceless (which is
the case for the spin models considered in the main text).
We employ a high-temperature expansion of $\rho_\beta \sim (I-\beta H)/\mathcal{N}_\mathcal{H} + \mathcal{O}(\beta^2)$. To first order in $\beta$ we find
\begin{eqnarray}
O_S[\rho_\beta](t) 
&=&O(t)-\frac{\beta}{2\mathcal{N}_\mathcal{H}}(\mathrm{Tr}(HW(t)VW(t)V)
\nonumber \\
&+&\mathrm{Tr}(HVW(t)VW(t)))+\mathcal{O}(\beta^2).\label{eq:OS}
\end{eqnarray}
We now introduce the statistical correlations
\begin{equation}
\tilde C(t) = \overline{\langle W(t) \rangle_{u,k_0} \langle VW(t)V \rangle_{u,k_0} \langle H \rangle_{u,k_0} }, \nonumber
\end{equation}
with $u$ global random unitaries of the CUE and $\langle A \rangle_{u,k_0}=\bra{\psi_u}A \ket{\psi_u}$. Using the $3$-design properties~\cite{Collins2006} of the CUE, 
we obtain 
\begin{eqnarray}
\tilde C(t) = c' \sum_{\tau \in S_3} \mathrm{Tr}(\tau[W(t) \otimes V W(t)V \otimes H]),\nonumber
\end{eqnarray}
with $c'=[\mathcal{N}_\mathcal{H}(\mathcal{N}_\mathcal{H}+1)(\mathcal{N}_\mathcal{H}+2)]^{-1}$.
This can be rewritten as 
\begin{eqnarray}
\tilde C(t) = c' \left(\mathrm{Tr}(HW(t)VW(t)V)+\mathrm{Tr}(HVW(t)VW(t))\right).\label{eq:Ctilde}\nonumber
\end{eqnarray}
Using Eqs.~\eqref{eq:OS}-\eqref{eq:Ctilde}, and $O(t)=\tilde O^\mathrm{(G)}(t)$, we obtain
\begin{equation}
O_S[\rho_\beta](t) =O(t)-\frac{\beta}{2c'\mathcal{N}_\mathcal{H}} \tilde C(t), 
\end{equation}
which shows that $O_S[\rho_\beta](t)$ can be accessed by measuring separately (i) the $T=\infty$ value $O(t)$ as described in the main text, and (ii) the additional correlations $\tilde C(t)$. This shows
that thermal OTOCs are experimentally accessible by measuring statistical correlations. The additional requirement compared to the measurement of $O(t)$ is the measurement of the energy $\langle H \rangle_{u,k_0}$ of random initial sates. 
\input{ProbingScrambling.bbl}
\end{document}

%% file: ProbingScrambling.bbl
%